\documentclass[a4paper]{article}
\usepackage{indentfirst}
\usepackage{hyperref}
\usepackage{multirow}
\usepackage{latexsym,bm,amsmath,amssymb}
\usepackage{graphicx}
\usepackage{epsfig}
\usepackage{subfigure}
\usepackage{cite}
\usepackage{color}
\usepackage[top=1in, bottom=1in, left=1.2in, right=1.2in]{geometry}
\usepackage{appendix}
\usepackage{ulem}
\usepackage{slashed}

\newcommand{\bea}{\begin{eqnarray}}
\newcommand{\eea}{\end{eqnarray}}
\newcommand{\tabincell}[2]{\begin{tabular}{@{}#1@{}}#2\end{tabular}}
\begin{document}
\title{New Insights on Low Energy $\pi N$ Scattering Amplitudes}
\author
{
Yu-Fei Wang$^{1}$, De-Liang~Yao$^{2}$, Han-Qing~Zheng$^{1,3}$
\vspace*{0.3cm} \\
$^{1}${\it Department of Physics and State Key Laboratory of Nuclear Physics and Technology,} \\
{\it Peking University,  Beijing 100871, China}\\
$^{2}${\it Instituto de F\'{\i}sica Corpuscular (centro mixto CSIC-UV), }\\
{\it Institutos de Investigaci\'{o}n de Paterna, } \\
{\it Apartado 22085, 46071, Valencia, Spain}\\
$^{3}${\it  Collaborative Innovation Center of Quantum Matter, Beijing 100871, China}
}
\maketitle
\begin{abstract}
The $S$- and $P$- wave phase shifts of low-energy pion-nucleon scatterings are analysed using Peking University representation, in which they are decomposed into various terms contributing either from poles or branch cuts. We estimate the left-hand cut contributions with the help of tree-level perturbative amplitudes derived in relativistic baryon chiral perturbation theory up to $\mathcal{O}(p^2)$. It is found that in $S_{11}$ and $P_{11}$ channels, contributions from known resonances and cuts are far from enough to saturate experimental phase shift data -- strongly indicating contributions from low lying poles undiscovered before, and we fully explore possible physics behind. On the other side, no serious disagreements are observed in the other channels.
\end{abstract}
\section{Introduction}
The pion-nucleon ($\pi N$) elastic scattering, as one of the most fundamental and important processes in nuclear or hadron physics, has been studied for decades~\cite{Bransden:1973,Hohler:1983}. However, there are still many open questions need to be attained more insights into. For instance, the low energy behavior of the $\pi N$ elastic scattering amplitude, the pion nucleon $\sigma$-term and the relevant intermediate resonances, e.g., $\Delta(1232)$, $N^*(1535)$ and $N^*(1440)$, have attracted sustained attentions, see, e.g., Refs.~\cite{Becher:2001hv,Lutz:2010,Ditsche:2012fv,AlarconA,Alarcon:2012kn,Chen:2012nx,Hoferichter:2015dsa,Yao:2016vbz}. The $N^*(1535)$ and $N^*(1440)$ are of particular interest. For $N^*(1535)$, the origin of its high mass and its large coupling to the $\eta N$ channel have been studied in the literature~\cite{Kaiser:1995cy,Nieves:2001wt}. As for $N^*(1440)$, its quark model interpretation and its coupling to $\sigma N$ channel are still not well understood~\cite{Krehl:1999km}. Furthermore, it may contain a two-pole structure~\cite{Arndt:1985vj}, and the corresponding $P_{11}$ channel may have strange branch cuts in the complex $s$ plane~\cite{Ceci:2011ae}. In this paper, we adopt another approach to study the low energy $\pi N$ scattering amplitudes. Peking University (PKU) representation~\cite{Xiao:2000kx,He:2002ut,Zheng:2003rw,Zhou:2004ms,Zhou:2006wm} is a model-independent method based on axiomatic $S$-matrix arguments. It has been successfully applied to investigate $\pi\pi$ and $\pi K$ scatterings and, in particular, corroborate the existences of $\sigma$ and $\kappa$ resonances~\cite{Xiao:2000kx,Zheng:2003rw}.  The use of PKU representation to study $\pi N$ scatterings may help us not only to enrich our knowledge of the amplitude structure but also to gain a fresh look at relevant physics in a much more rigorous manner.

The PKU representation factorizes the partial wave two-body elastic scattering $S$ matrix in the form~\cite{Zheng:2003rw}
\begin{equation}\label{PKU}
\begin{split}
S(s)&=\prod_b \frac{1-\text{i}\rho(s)\frac{s}{s-s_L} \sqrt{\frac{s_b-s_L}{s_R-s_b}}}
{1+\text{i}\rho(s)\frac{s}{s-s_L} \sqrt{\frac{s_b-s_L}{s_R-s_b}}}
\prod_v \frac{1+\text{i}\rho(s)\frac{s}{s-s_L} \sqrt{\frac{s_v'-s_L}{s_R-s_v'}}}
{1-\text{i}\rho(s)\frac{s}{s-s_L} \sqrt{\frac{s_v'-s_L}{s_R-s_v'}}}\\
&\times\prod_r \frac{M^2_r-s+\text{i}\rho(s)sG_r}{M^2_r-s-\text{i}\rho(s)sG_r}
e^{2\text{i}\rho(s)f(s)}\ \mbox{, }
\end{split}
\end{equation}
where the functions in resonance terms read
\begin{align}
&M^2_r=\text{Re}[z_r]+\text{Im}[z_r]\frac{\text{Im}[\sqrt{(z_r-s_R)(z_r-s_L)}]}{\text{Re}[\sqrt{(z_r-s_R)(z_r-s_L)}]}\ \mbox{, }\\
&G_r=\frac{\text{Im}[z_r]}{\text{Re}[\sqrt{(z_r-s_R)(z_r-s_L)}]}\ \mbox{, }
\end{align}
and the kinematic factor is defined by
\begin{equation}\label{rhodef}
\rho(s)=\frac{\sqrt{s-s_L}\sqrt{s-s_R}}{s}
\end{equation}
with $s_L=(m_1-m_2)^2$ and $s_R=(m_1+m_2)^2$.  Here the masses of the two scattering particles are labeled by $m_1$ and $m_2$. Furthermore, $s_b,\ s_v'$ and $z_r$ denote bound state poles (on the real axis below threshold of the first Riemann sheet), virtual state poles (on the real axis below threshold of the second Riemann sheet) and resonances (on the second Riemann sheet off the real axis), respectively. Lastly, the exponential term in Eq.~(\ref{PKU}) is named as background term since it contains no poles. Actually, the background term carries the information of left-hand cuts (\textit{l.h.c.}s) and right-hand inelastic cut {(\textit{r.h.i.c.})} above inelastic thresholds, and it satisfies a dispersion relation,
\begin{equation}\label{fdisper}
f(s)=\frac{s}{2\pi\text{i}}\int_Lds'\frac{\text{disc}f(s')}{(s'-s)s'}+\frac{s}{2\pi\text{i}}\int_{\text{R}'}ds'\frac{\text{disc}f(s')}{(s'-s)s'} \ \mbox{, }
\end{equation}
where L and R$'$ denote the \textit{l.h.c.}s and \textit{r.h.i.c.} respectively, and disc stands for the discontinuity of the function $f(s)$ along the cuts. To obtain the background function $f(s)$ in Eq.~(\ref{fdisper}), a first-order subtraction has been performed at the point $s_0=0$, at which the subtraction constant term $f(s_0)$ vanishes, i.e. $f(0)=0$, as pointed out in Ref~\cite{Zhou:2006wm}. It should be emphasized that resonances other than the second sheet ones are not presented in the resonance terms in Eq.~(\ref{PKU}), rather, their contributions are hidden in the \textit{r.h.i.c.} integral of Eq.~(\ref{fdisper})~\cite{Zhou:2006wm}.

PKU representation is derived based on first principles of $S$-matrix theory, thus, in principle it is rigorous and universal for two-body elastic scatterings.\footnote{It is only confined to the situation of elastic scatterings. For coupled channel situation, a production representation is not established, see Ref.~\cite{Xiao:2001pt}. } Besides, the production representation equipped itself with additive phase shifts stemming from different contributions, which makes the analysis of phase shifts clear and convenient, and enables one even to find out hidden contributions. The phase shifts given by PKU representation are sensitive to (not too) distant poles, letting one determine pole positions rather accurately. Moreover, each phase shift contribution has a definite sign: bound states always give negative contributions, while virtual states and resonances always give positive contributions.\footnote{Actually, the PKU representation method is the quantum field theory correspondence of Ning Hu representation in quantum mechanical scattering theory, see Ref.~\cite{Hu:1948zz}. } Furthermore, the \textit{l.h.c.}s would give negative phase shifts\footnote{This lacks of {rigorous} mathematical prove, but is correct empirically, and is to be discussed later at tree level. There exists a corresponding prove at the level of quantum mechanical scattering theory under some assumptions, see Ref.~\cite{Regge:1958ft}. }. The observation that the \textit{l.h.c.}s give a large and negative contribution is crucial to firmly establish the very existence of the $\sigma$ meson {($f_0(500)$)} in Ref.~\cite{Xiao:2000kx}, and the $\kappa$ resonance ($K^*(800)$) in Ref.~\cite{Zheng:2003rw}. The essence of PKU representation is not to directly unitarize the amplitude itself, rather, it unitarizes the \textit{l.h.c.}s of the perturbative amplitude and hence hazardous spurious poles can be avoided. Specific example concerning the advantage of PKU representation, compared to some conventional unitarization approaches (like Pad\'{e} approximation), can be found in Ref.~\cite{Qin:2002hk}.

In this paper, PKU representation is employed to study the $S$- and $P$- wave channels of the $\pi N$ elastic scattering. On the one hand, the various relevant poles are incorporated as inputs or determined by fit. {On the other hand}, the contribution of \textit{l.h.c.}s is deduced with the help of chiral perturbative amplitudes of $\mathcal{O}(p^2)$ at tree level, derived in a relativistic baryon chiral perturbation theory (BChPT)~\cite{Bernard:1995dp}. In the following section \ref{cal}, the basic formulae relevant to the calculation are shown. In section \ref{res} the numerical results are presented and discussed. Finally section \ref{con} contains conclusions and outlook of this paper. The tree amplitudes up to $\mathcal{O}(p^2)$ and their partial-wave projection are relegated to Appendices~\ref{app:ff} and~\ref{app:pw}, respectively. The major uncertainties of the estimation of the background integral are compiled in Appendix~\ref{app:uncer}.

\section{Theoretical framework}\label{cal}
\subsection{Left-hand cut contributions implied by BChPT at tree-level}
In the $SU(2)$ isospin limit, the $\pi N$ Lagrangians relevant to the calculation up to $\mathcal{O}(p^2)$ are~\cite{Fettes:2000gb}:
\begin{equation}\label{p1la}
\mathcal{L}_{\pi N}^{(1)}=\bar N\left(\text{i}\slashed{D}-M+\frac{1}{2}g\slashed{u}\gamma^5\right)N\ \mbox{, }
\end{equation}
\begin{equation}\label{p2la}
\begin{split}
\mathcal{L}_{\pi N}^{(2)}&=c_1\langle\chi_+\rangle\bar NN-\frac{c_2}{4M^2}\langle u^\mu u^\nu\rangle(\bar N D_\mu D_\nu N+\text{h.c.})\\
&+\frac{c_3}{2}\langle u^\mu u_\mu\rangle\bar NN-\frac{c_4}{4}\bar N \gamma^\mu\gamma^\nu \big[u^\mu,u^\nu\big]N\ \mbox{, }
\end{split}
\end{equation}
with $M$ being the mass of the nucleon, $g$ being the axial current coupling constant and $c_i(i=1,2,3,4)$ the $\mathcal{O}(p^2)$ coupling constants. The ``$\langle\cdots\rangle$'' denotes the matrix tracing in isospin space.  The pion fields are compiled in
\begin{equation}\label{pif}
u(x)=\exp\left(\frac{\text{i}\pi^a\tau^a}{2F}\right)\ \mbox{, }
\end{equation}
with $F$ being the pion decay constant in the chiral limit and $\tau^a$ standing for Pauli matrices. The chiral building blocks in Eqs.~(\ref{p1la}) and (\ref{p2la}) are as follows:
\[
\begin{split}
&D_\mu=\partial_\mu+\Gamma_\mu\ \mbox{, }\\
&\Gamma_\mu=\frac{1}{2}\big[u^\dagger(\partial_\mu-\text{i}r_\mu)u+u(\partial_\mu-\text{i}l_\mu)u^\dagger\big]\ \mbox{, }\\
&u_\mu=\text{i}\big[u^\dagger(\partial_\mu-\text{i}r_\mu)u-u(\partial_\mu-\text{i}l_\mu)u^\dagger\big]\ \mbox{, }\\
&\chi_{+}=u^\dagger\chi u^\dagger+ u\chi^\dagger u\ \mbox{, }\\
&\chi=2B_0(s+i p)\ \mbox{, }
\end{split}
\]
During the procedure of calculation one needs to set $2B_0s\to2 B_0m_q\equiv m^2$ with $m$ being the pion mass, while the other sources ($l_\mu,\ r_\mu$ and $p$) are switched off. To obtain the amplitudes with definite isospin $I$ and angular momentum $J$, one should decompose the isospin structure and then perform partial wave projection. For isospin decomposition,
\begin{equation}
T(\pi^a+N_{\text{i}}\to \pi^{a'}+N_{\text{f}})=\chi_{\text{f}}^\dagger\left(\delta^{a'a}T^S+\frac{1}{2}\big[\tau^{a'},\tau^a\big]T^A\right)\chi_{\text{i}}\ \mbox{, }
\end{equation}
where $\chi_{\text{i}}$ and $\chi_{\text{f}}$ are isospinors of initial and final nucleon states, respectively. Then the amplitudes with isospins $I=\frac{1}{2},\frac{3}{2}$ can be written as
\begin{align}
&T^{I=1/2}=T^S+2T^A\ ,\\
&T^{I=3/2}=T^S-T^A\ \mbox{. }
\end{align}
Further, the Lorentz structure of the above isospin amplitudes reads
\begin{equation}
T^I=\bar u(p',s')\Big[A^I(s,t)+\frac{1}{2}(\slashed{q}+\slashed{q}')B^I(s,t)\Big]u(p,s)\ \mbox{, }
\end{equation}
where $s,t$ are Mandelstam variables, and $q$ and $q'$ are the $4$-momenta of initial and final states of pions, respectively. The tree-level $A^I$ and $B^I$ up to $\mathcal{O}(p^2)$ are listed in Appendix.~\ref{app:ff}. The helicity amplitudes in the centre of mass frame can be expressed in terms of functions $A^I$ and $B^I$ as
\begin{align}
&T_{++}^I=(\frac{1+z_s}{2})^{\frac{1}{2}}[2MA^I(s,t)+(s-m^2-M^2)B^I(s,t)]\ \mbox{, }\\
&T_{+-}^I=-(\frac{1-z_s}{2})^{\frac{1}{2}}s^{-\frac{1}{2}}[(s-m^2+M^2)A^I(s,t)+M(s+m^2-M^2)B^I(s,t)]\ \mbox{, }
\end{align}
where the subscripts ``$\pm$''  are abbreviations of helicity $h=\pm1/2$. Moreover, the first and second subscripts correspond to the helicities of the initial and final nucleon states, respectively. $z_s$ is defined as the cosine of the scattering angle. The partial wave projection formulae are given by
\begin{align}
&T_{++}^{I,J}=\frac{1}{32\pi}\int_{-1}^1 dz_s T_{++}^I(s,t(s,z_s)) d^J_{-1/2,-1/2}(z_s)\ \mbox{, }\\
&T_{+-}^{I,J}=\frac{1}{32\pi}\int_{-1}^1 dz_s T_{+-}^I(s,t(s,z_s)) d^J_{1/2,-1/2}(z_s)\ \mbox{, }
\end{align}
with $d^J$ to be the Wigner $D$-matrix. To be specific, the six $S$- and $P$- wave amplitudes (in $L_{2I\ 2J}$ convention) can be represented in terms of helicity amplitudes as follows:
\begin{align}
&T(S_{11})=T_{++}^{1/2,1/2}+T_{+-}^{1/2,1/2}\ \mbox{, }\nonumber\\
&T(S_{31})=T_{++}^{3/2,1/2}+T_{+-}^{3/2,1/2}\ \mbox{, }\nonumber\\
&T(P_{11})=T_{++}^{1/2,1/2}-T_{+-}^{1/2,1/2}\ \mbox{, }\nonumber\\
&T(P_{31})=T_{++}^{3/2,1/2}-T_{+-}^{3/2,1/2}\ \mbox{, }\nonumber\\
&T(P_{13})=T_{++}^{1/2,3/2}+T_{+-}^{1/2,3/2}\ \mbox{, }\nonumber\\
&T(P_{33})=T_{++}^{3/2,3/2}+T_{+-}^{3/2,3/2}\ \mbox{. }
\end{align}
The explicit expressions of the partial-wave helicity amplitudes can be found in Appendix~\ref{app:pw}. Eventually, the discontinuity of function $f$ can be deduced through (the symbols of the channels are omitted)
\begin{align}
&\text{disc}\big[f(s)\big]=\text{disc}\left[\frac{\ln S(s)}{2i\rho(s)}\right]\ ,\nonumber\\
&S(s)=1+2i\rho(s)T(s)\ ,
\end{align}
where $T(s)$ is perturbatively calculated here, and function $f(s)$ can be obtained by using Eq.~(\ref{fdisper}).
At tree level, the left-hand cut structure for $T$ is quite simple: a kinematic cut $(-\infty,0]$ and a segment cut $[(M^2-m^2)^2/M^2,2m^2+M^2]$ due to the $u$-channel nucleon exchange. Note that $\rho(s)$ has an extra branch-cut $(-\infty,(M-m)^2]$ by definition in Eq.~\eqref{rhodef}, so the cuts of $S(s)$ should be $(-\infty,(M-m)^2]$ and $[(M^2-m^2)^2/M^2,2m^2+M^2]$\footnote{The full analytic structure of the \textit{l.h.c.}s can be found in Ref.~\cite{Kennedy:1961}. At $\mathcal{O}(p^2)$ level, there is no circular cut. }, thus the background function is
\begin{equation}\label{fsint}
f(s)=-\frac{s}{\pi}\int_{s_{c}}^{(M-m)^2} \frac{\ln|S(w)|dw}{2\rho(w)w(w-s)}+\frac{s}{\pi}\int_{(M^2-m^2)^2/M^2}^{2m^2+M^2} \frac{\text{Arg}[S(w)]dw}{2iw\rho(w)(w-s)}\ \mbox{. }
\end{equation}
From Eq.~\eqref{fsint} it is found that the dispersion integral contains a logarithmic term and once subtraction, which significantly suppresses the bad behavior of perturbation theory in high energy region -- even if the integral domain of the first term is chosen to be $(-\infty,(M-m)^2]$, the integral still converges. That property guarantees the results to be insensitive to high energy contributions. However, perturbative calculations would inevitably become invalid when $w$ is too large, so one has to assign the integral domain a cut-off parameter $s_{c}$. In principle the exact value of $s_c$ is unknown, and may be fixed by fitting to the data if one tentatively regards Eq.~\ref{fsint} as a parameterization of the left-hand cut. On the other side, it is also educative to chose $s_c$ to be at the boundary of perturbation theory convergence region, in order to evaluate the contributions from where perturbation theory is valid. Of course, meaningful physical outputs should be immune from such an ambiguity of the left-hand cut integral.

Actually the contribution from the $u$-channel cut is numerically very small\footnote{This is due to that the near threshold $u$ channel exchange $1/(u-M^2)$ can be approximately represented as a contact interaction, which leaves no left-hand cut at all. }, hence the dominant contribution is from the first term of Eq.~\eqref{fsint}, which is always negative in physical region ($s>(M+m)^2$), since $\rho(w)w(w-s)>0$ when $w<(M-m)^2$ and $|S|>1$ in perturbation theory. Besides, the \textit{r.h.i.c.} is not considered for the moment, since the energy region to be analyzed (from the $\pi N$ threshold to $1.16$ GeV) is below the inelastic threshold, where the \textit{r.h.i.c.} contribution is empirically small. However, this topic is to be discussed further in Section \ref{rhccal}.
\subsection{Known-pole contributions estimated from experiments}
The above discussions are devoted to the estimation of the cut contributions from chiral perturbative amplitudes. In practice, one also needs to take into account the pole contributions. The known poles (nucleon bound state and above-threshold resonances), listed in table~\ref{tab:poles}~\cite{Anisovich:2011fc}, are under our consideration.
\begin{table}[htbp]
\begin{center}
 \begin{tabular}  {| c | c | c ||}
  \hline
  Channels  & $I(J^P)$  & Poles\\
  \hline
  $S_{11}$ &$\frac{1}{2}({\frac{1}{2}}^-)$ & $N^*(1535), N^*(1650), N^*(1895)$\\
  \hline
  $S_{31}$ &$\frac{3}{2}({\frac{1}{2}}^-)$ & $\Delta(1620), \Delta(1900)$\\
  \hline
  $P_{11}$ &$\frac{1}{2}({\frac{1}{2}}^+)$ & $N, N^*(1440), N^*(1710), N^*(1880)$\\
  \hline
  $P_{31}$ &$\frac{3}{2}({\frac{1}{2}}^+)$ & $\Delta(1910)$\\
  \hline
  $P_{13}$ &$\frac{1}{2}({\frac{3}{2}}^+)$ & $N^*(1720), N^*(1900)$\\
  \hline
  $P_{33}$ &$\frac{3}{2}({\frac{3}{2}}^+)$ & $\Delta(1232),\Delta(1600),\Delta(1920)$\\
  \hline
 \end{tabular}\\
 \caption{Intermediate poles added in each channel. }\label{tab:poles}
\end{center}
\end{table}
However, the PKU method can only deal with poles on the first and second Riemann sheets, namely, the poles located on the third Riemann sheet given by experiments
\[
\sqrt{s}^{\text{III}}=M_{\text{pole}}-\frac{i}{2}(\Gamma_{\text{inelastic}}+\Gamma_{\pi N})
\]
cannot be used directly\footnote{The information of the poles on third or higher sheets is hidden in the inelastic cut in Eq.~(\ref{fdisper}) and their contributions are rather indirect~\cite{Zhou:2006wm}. }. Under narrow width approximation, the second sheet poles (usually called shadow poles) may be estimated by
\begin{equation}\label{shadowpole}
\sqrt{s}^{\text{II}}=M_{\text{pole}}-\frac{i}{2}(\Gamma_{\text{inelastic}}-\Gamma_{\pi N})\ \mbox{. }
\end{equation}
With the preparations made in this section, we proceed with the numerical study in the next section.
\section{Numerical results and discussions}\label{res}
\subsection{Prelude: a $K$-matrix fit and spurious poles}
\begin{figure}[htbp]
\center
\subfigure[]{
\label{com:subfig:S11Kfit}
\scalebox{1.0}[1.0]{\includegraphics[width=0.4\textwidth]{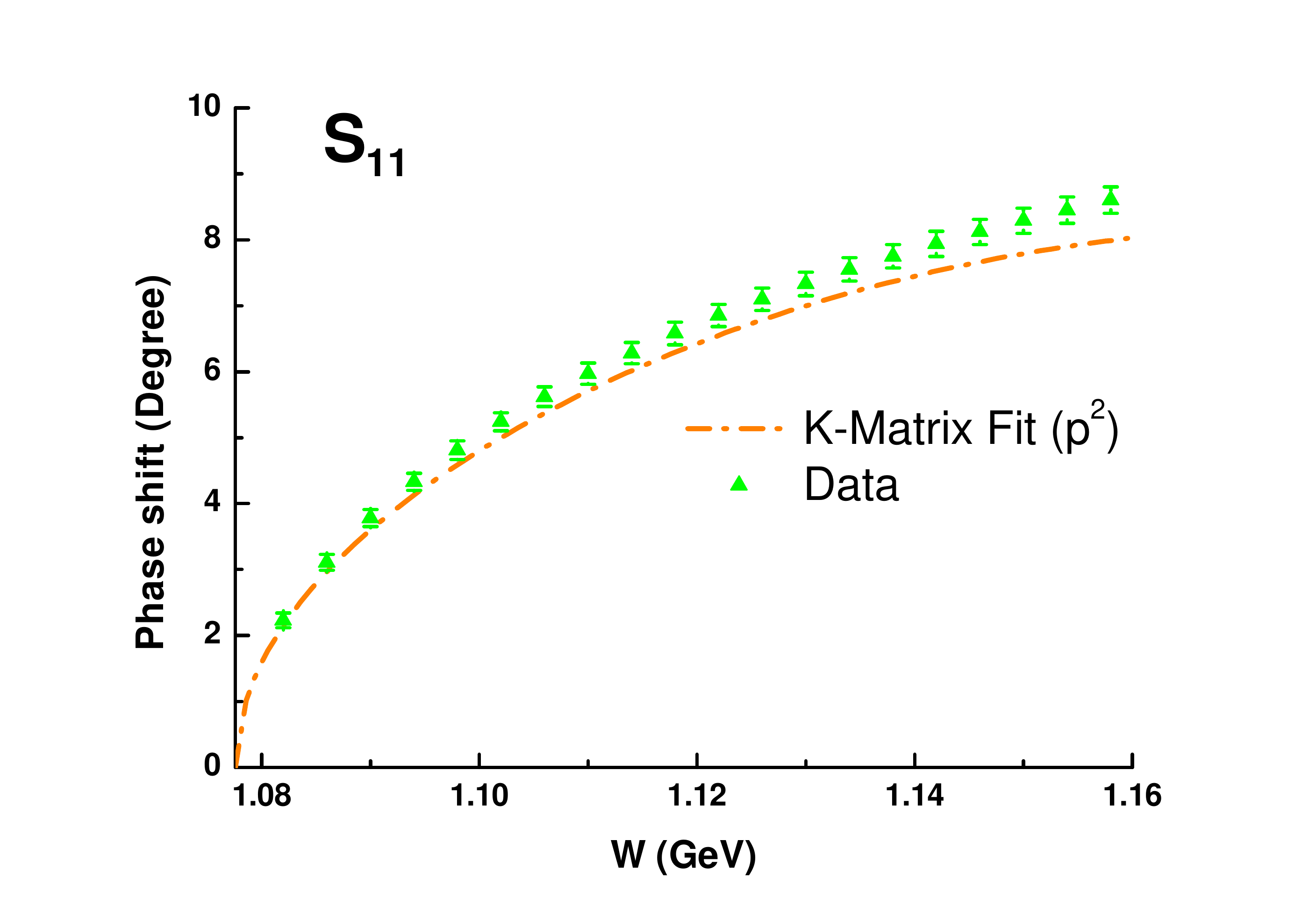}}}
\subfigure[]{
\label{com:subfig:S31Kfit}
\scalebox{1.0}[1.0]{\includegraphics[width=0.4\textwidth]{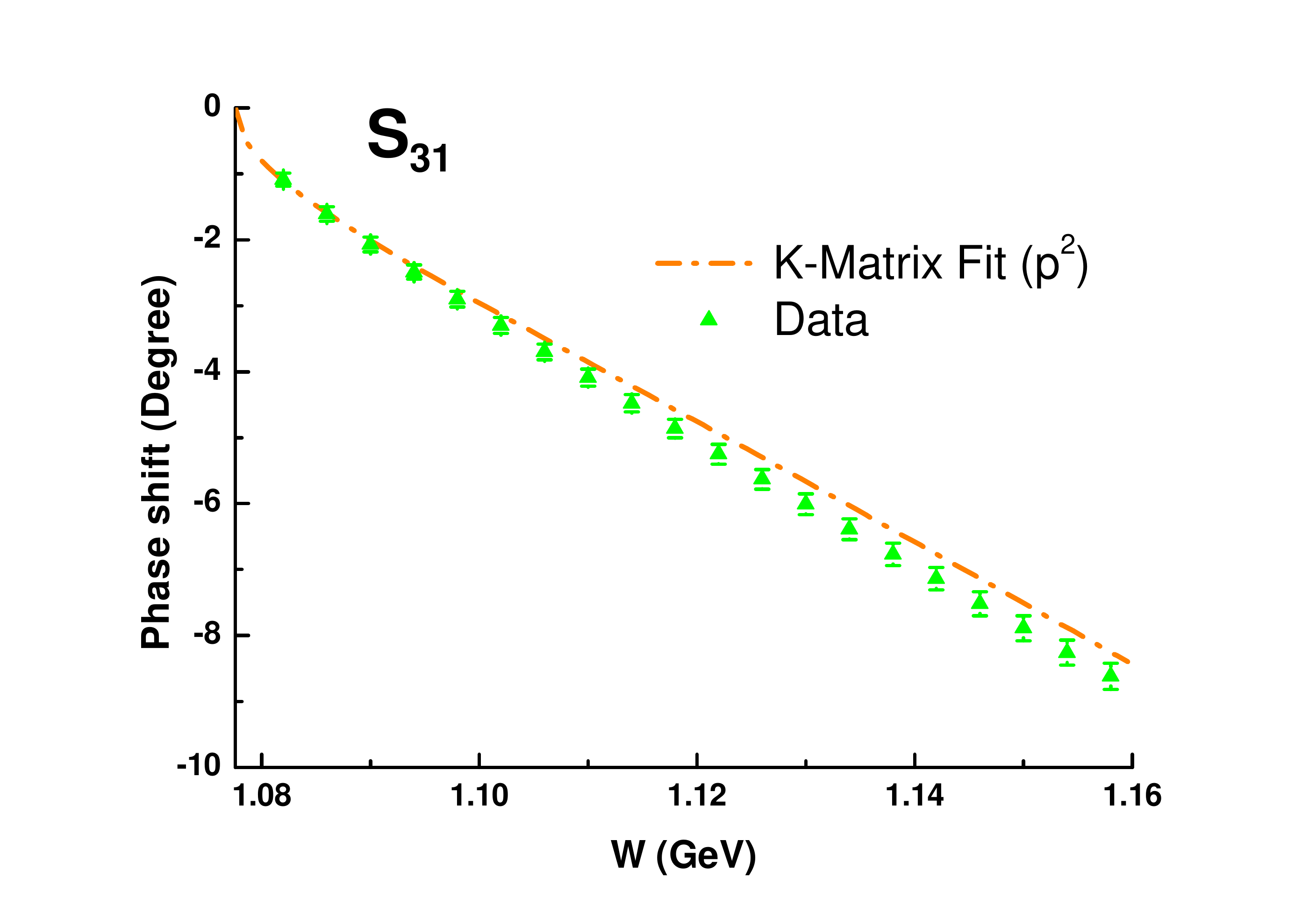}}}
\subfigure[]{
\label{com:subfig:P11Kfit}
\scalebox{1.0}[1.0]{\includegraphics[width=0.4\textwidth]{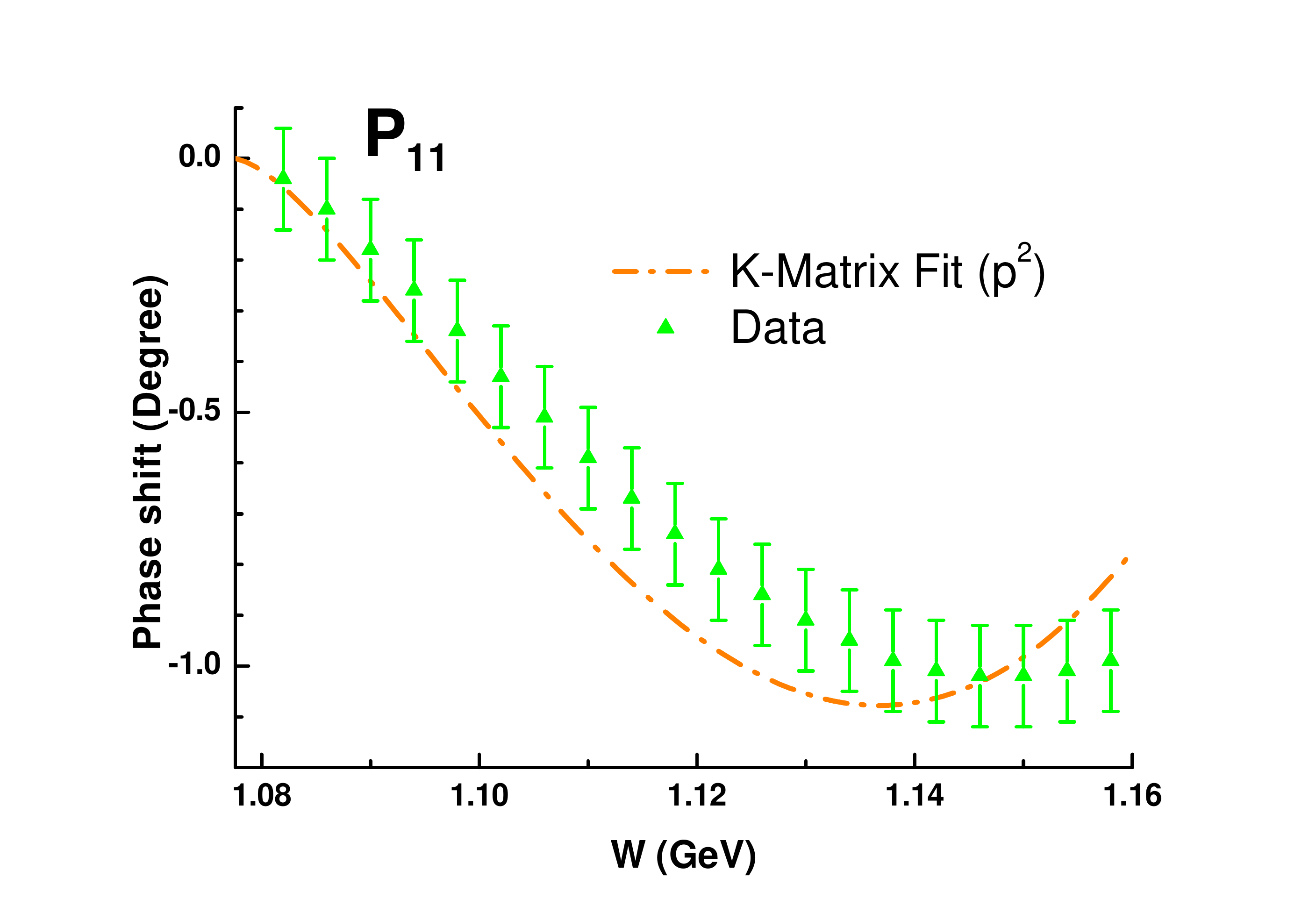}}}
\subfigure[]{
\label{com:subfig:P31Kfit}
\scalebox{1.0}[1.0]{\includegraphics[width=0.4\textwidth]{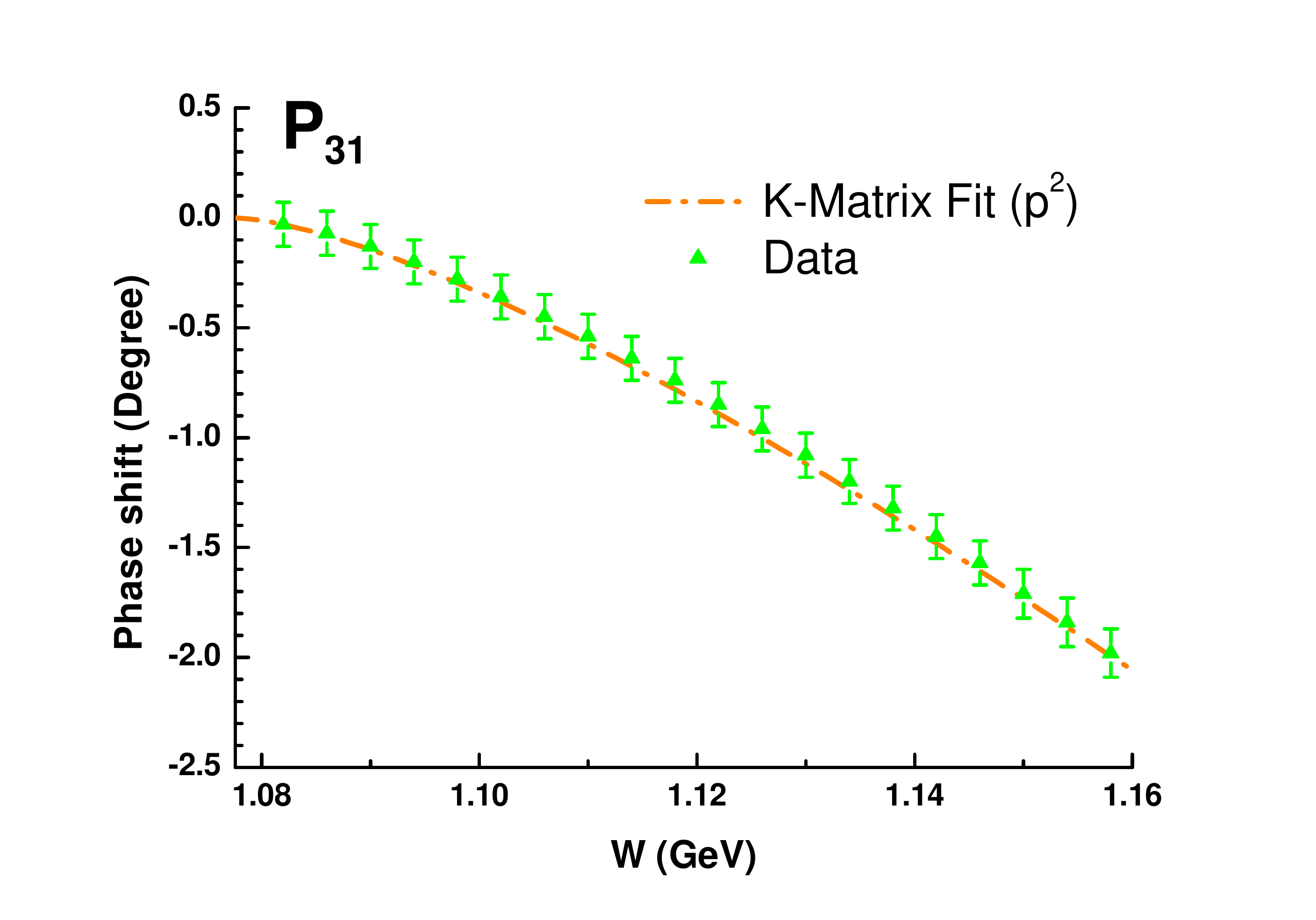}}}
\subfigure[]{
\label{com:subfig:P13Kfit}
\scalebox{1.0}[1.0]{\includegraphics[width=0.4\textwidth]{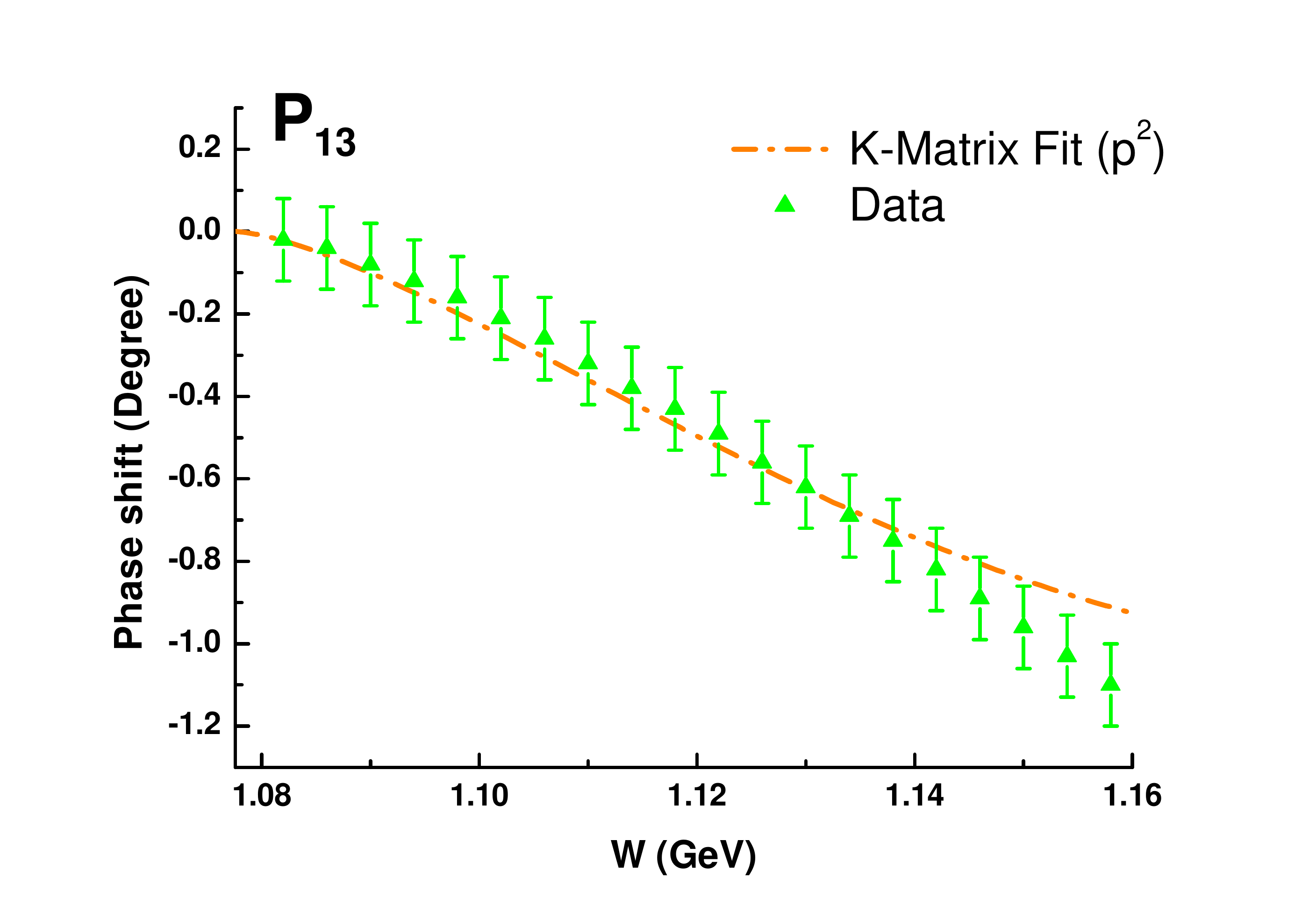}}}
\caption{$\mathcal{O}(p^2)$ $K$-matrix fit results of five channels. }\label{fig:Kfit}
\end{figure}

Here we use $K$-matrix method to determine the coupling constants $c_i$. The influence of different choices of $c_i$ parameters is mild {\color{red} as} discussed in Appendix.~\ref{app:ci}. In our numerical computation, values of the masses and $\mathcal{O}(p^1)$ coupling constants, see Eq.~\eqref{p1la}, are taken from Ref.~\cite{Patrignani:2016xqp}:  $M=0.9383~\text{GeV}$, $m=0.1396~\text{GeV}$, $F=0.0924~\text{GeV}$ and $g=1.267$. A $K$-matrix fit is performed to the data of $S_{11}$, $S_{31}$, $P_{11}$, $P_{31}$ and $P_{13}$ channels\footnote{The $P_{33}$ channel cannot fit to the data well at tree level without explicit $\Delta(1232)$ field, hence is excluded. } provided by George Washington University (GWU) group~\cite{SAID}. Unfortunately, the uncertainties of data are not provided by GWU group. Therefore, we assign errors to the data in the same way as done in Ref.~\cite{Chen:2012nx}. The tree-level $K$-matrix formula is
\begin{equation}\label{Kfit}
\begin{split}
&T_K=\frac{T}{1-i\rho\, T}\ \mbox{,}\\
&\delta_K=\arctan\big[\rho\, T\big]\ \mbox{. }
\end{split}
\end{equation}
We fit $20$ data points (corresponding to $W\equiv\sqrt{s}\in [1.0776,1.1600]$ GeV) in each channel.
The fitted values of parameters are
\begin{equation}\label{civalue}
c_1=-0.841\ \text{GeV}^{-1},\,c_2=1.170\ \text{GeV}^{-1},\,c_3=-2.618\ \text{GeV}^{-1},\,c_4=1.677\ \text{GeV}^{-1}\ ,
\end{equation}
with the fit quality $\chi^2/\text{d.o.f}=1.850$.
The fit results are plotted in Fig.~\ref{fig:Kfit}.

\begin{figure}[htbp]
\centering
\includegraphics[width=0.6\textwidth]{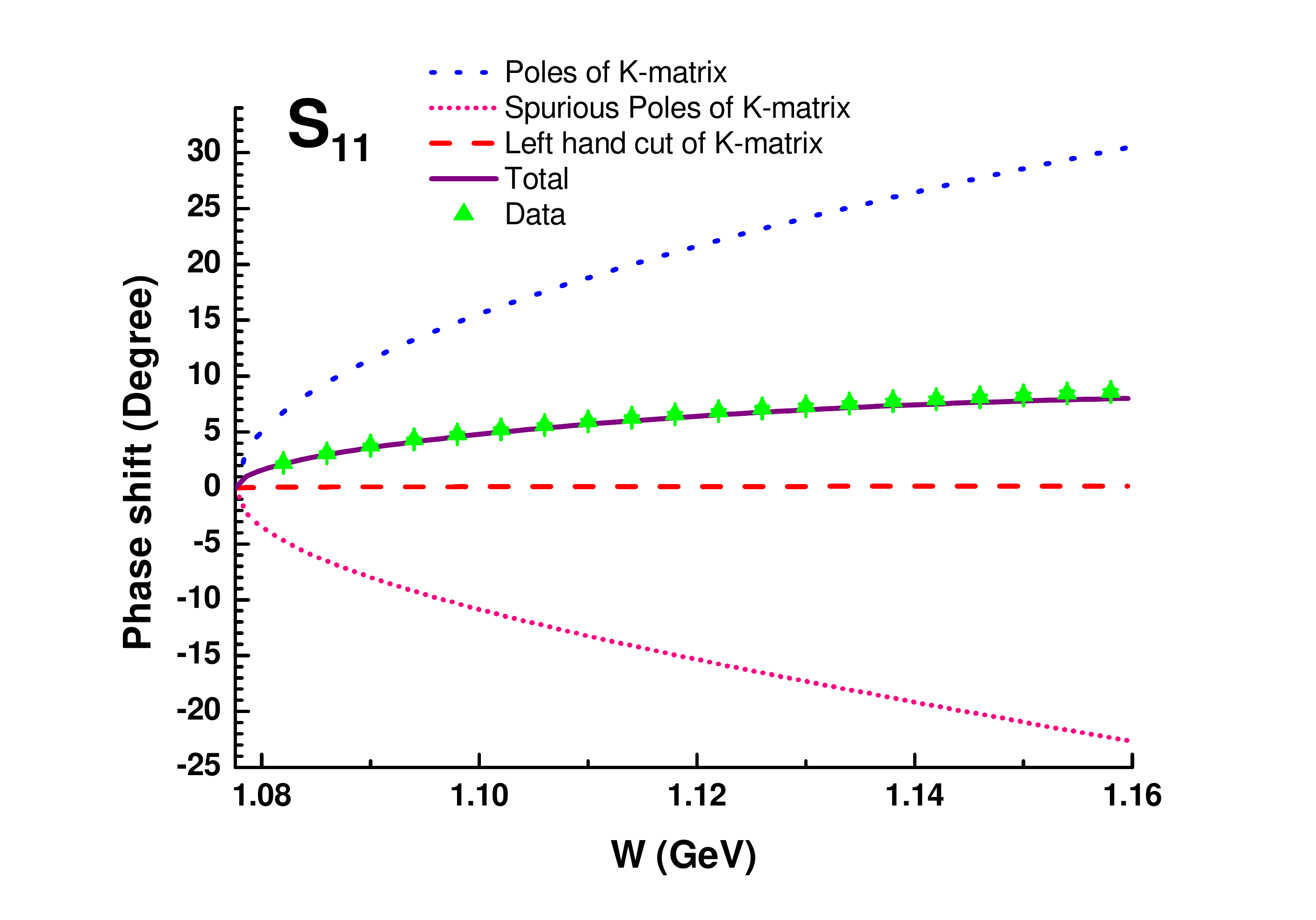}
\caption{PKU representation of the phase shift of $S_{11}$ channel based on $K$-matrix fit. }
\label{fig:S11p2K}
\end{figure}

Since $K$-matrix method gives the unitarized amplitudes (see Eq.~\eqref{Kfit}), it seems that poles can also be extracted. For instance, in $S_{11}$ channel, the poles given by $K$-matrix are listed in Table.~\ref{tab:S11K}.
\begin{table}[htbp]
\begin{center}
 \begin{tabular}  {| c | c ||}
  \hline
  Resonances (GeV) & spurious poles (GeV)\\
  \hline
  \tabincell{c}{$0.954-0.265 i$\\$2.254-0.067 i$} & \tabincell{c}{$0.733+0.089 i$\\$1.431+0.253 i$}\\
  \hline
 \end{tabular}\\
 \caption{The pole positions given by $K$-matrix method of $S_{11}$ channel. }\label{tab:S11K}
\end{center}
\end{table}
Firstly we discover that the $K$-matrix gives a near-threshold resonance locating at $0.954-0.265 i$ GeV, but the $K$-matrix also generates some poles on the first sheet off the real axis, which are called spurious poles\footnote{The spurious poles satisfy the equation $1-i\rho\, T=0$, and always give negative phase shifts in the convention of PKU representation. }, and have already been discussed in Ref.~\cite{Zheng:2003rw}. Those poles violate causality and hence are not allowed. Due to the existence of nearby spurious poles, the other poles given by $K$-matrix become unauthentic. If one uses PKU representation to separate the contributions in $K$-matrix, it is found that the ``good'' description of data rooted in the cancellation between large contributions from resonances and spurious poles, leaving a vigorously suppressed left-hand cut contribution, see Fig.~\ref{fig:S11p2K}.

\begin{figure}[htbp]
\centering
\includegraphics[width=0.6\textwidth]{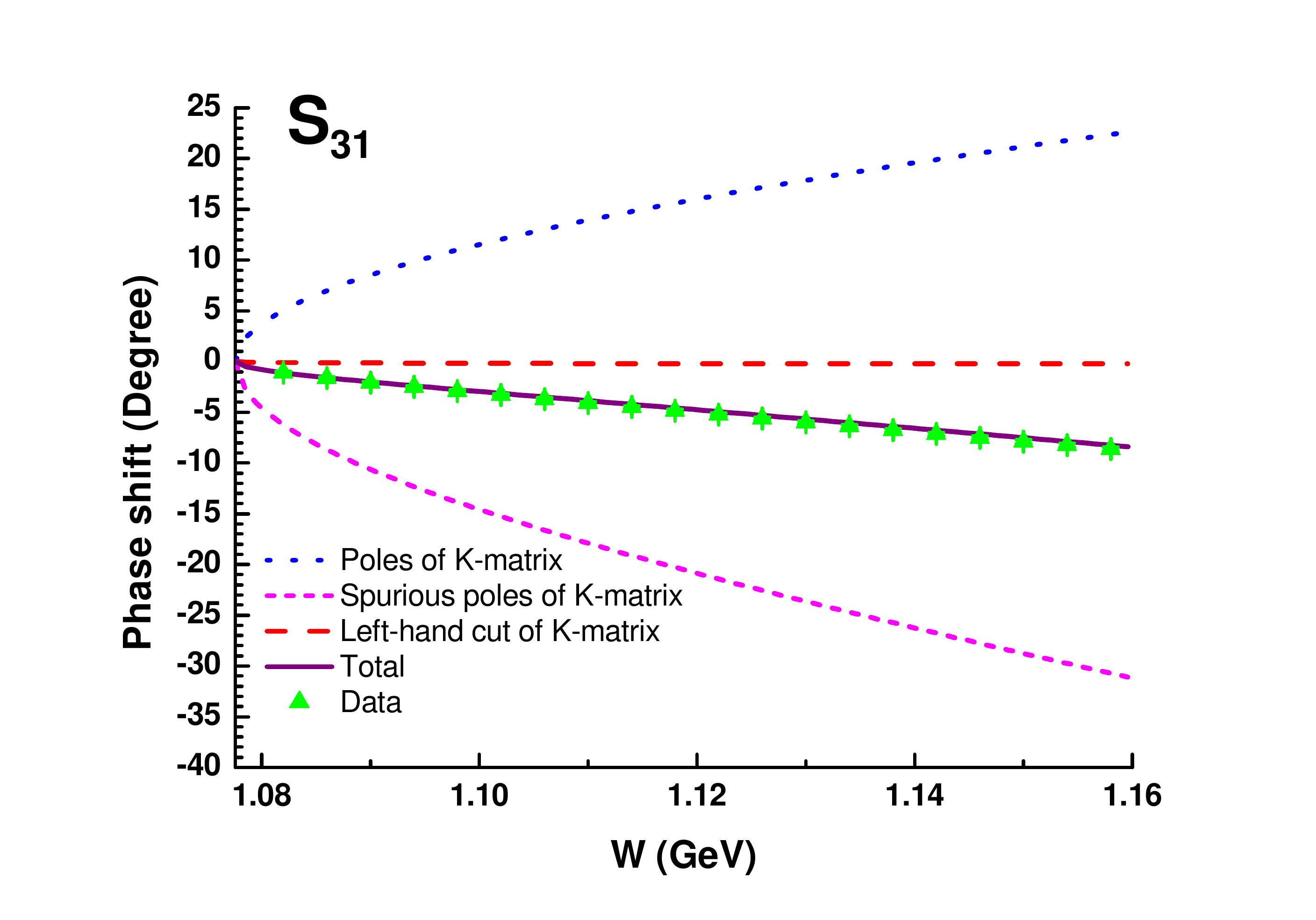}
\caption{PKU representation of the phase shift of $S_{31}$ channel based on $K$-matrix fit. }
\label{fig:S31p2K}
\end{figure}

In fact $K$-matrix amplitude in other channels may behave even worse. For example, in $S_{31}$ channel the $K$-matrix amplitude generates two bound states as well as two spurious poles, see Table.~\ref{tab:S31K}.
\begin{table}[htbp]
\begin{center}
 \begin{tabular}  {| c | c | c ||}
  \hline
  Bound states (GeV)  & spurious poles (GeV) & Resonances (GeV)\\
  \hline
  \tabincell{c}{$0.959$\\$0.916$} & \tabincell{c}{$0.647+0.064 i$\\$1.150+0.317 i$} & \tabincell{c}{$0.784-0.215 i$\\$0.938-0.0006 i$\\$1.626-0.142 i$}\\
  \hline
 \end{tabular}\\
 \caption{The pole positions given by $K$-matrix method of $S_{31}$ channel. }\label{tab:S31K}
\end{center}
\end{table}
The PKU representation analyses of the $K$-matrix amplitude in $S_{31}$ channel is plotted in Fig.~\ref{fig:S31p2K}.

In fact, the $K$-matrix method is constructed only relying on the right-hand unitarity, while PKU representation is more sophisticated since both analyticity and unitarity are maintained. From statements above, it is clear that the $K$-matrix approach is not at all admirable in determining distant poles, though we use it as a rough approximation to extract the $\mathcal{O}(p^2)$ coefficients. Therefore, it is pleasurable to use the strategy of PKU representation proposed in Refs~\cite{Zheng:2003rw,Zhou:2006wm,Xiao:2000kx,Zhou:2004ms} to tackle the spurious-pole problem.
\subsection{Pertinent PKU representation of phase shifts}\label{sec:PKU}
To apply PKU representation, one needs a value of the cut-off parameter $s_c$. As discussed before, we at first assess its value from the region where perturbation calculation works. That region can be obtained via different methods, e.g. through the unitarity bound (see Figure F.14 in Ref.\cite{Chen:2012nx}), or by assuming the validity of perturbation calculation till meeting the first resonance pole. Here we use the shadow pole location of $N^*(1440)$, which is the first resonance related to complicated couple-channel dynamics\footnote{In principle a complete calculation of the \textit{l.h.c.}s should contain $\Delta(1232)$ as a genuine state, but the \textit{l.h.c.} contribution from $t$ and $u$ channel $\Delta(1232)$ exchange is ignorable. }.
Since the chiral expansion is at the point $s=u=M^2+m^2,\ t=0$, we set the distance between $M^2+m^2$ and $N^*(1440)$ shadow pole location as the convergence radius $r$, and hence the cut-off on the left-hand cut is at $s_c=M^2+m^2-r\simeq -0.08$ GeV$^2$. In what follows, actually, it is seen that this choice of $s_c$ is reasonable in most channels; even if in quantitative analyses of $P_{11}$ channel the cut-off parameter should be tuned, the major physical outputs and conclusions are insensitive to its value.

\begin{figure}[htbp]
\center
\subfigure[]{
\label{p2PKU:subfig:S11}
\scalebox{1.2}[1.2]{\includegraphics[width=0.4\textwidth]{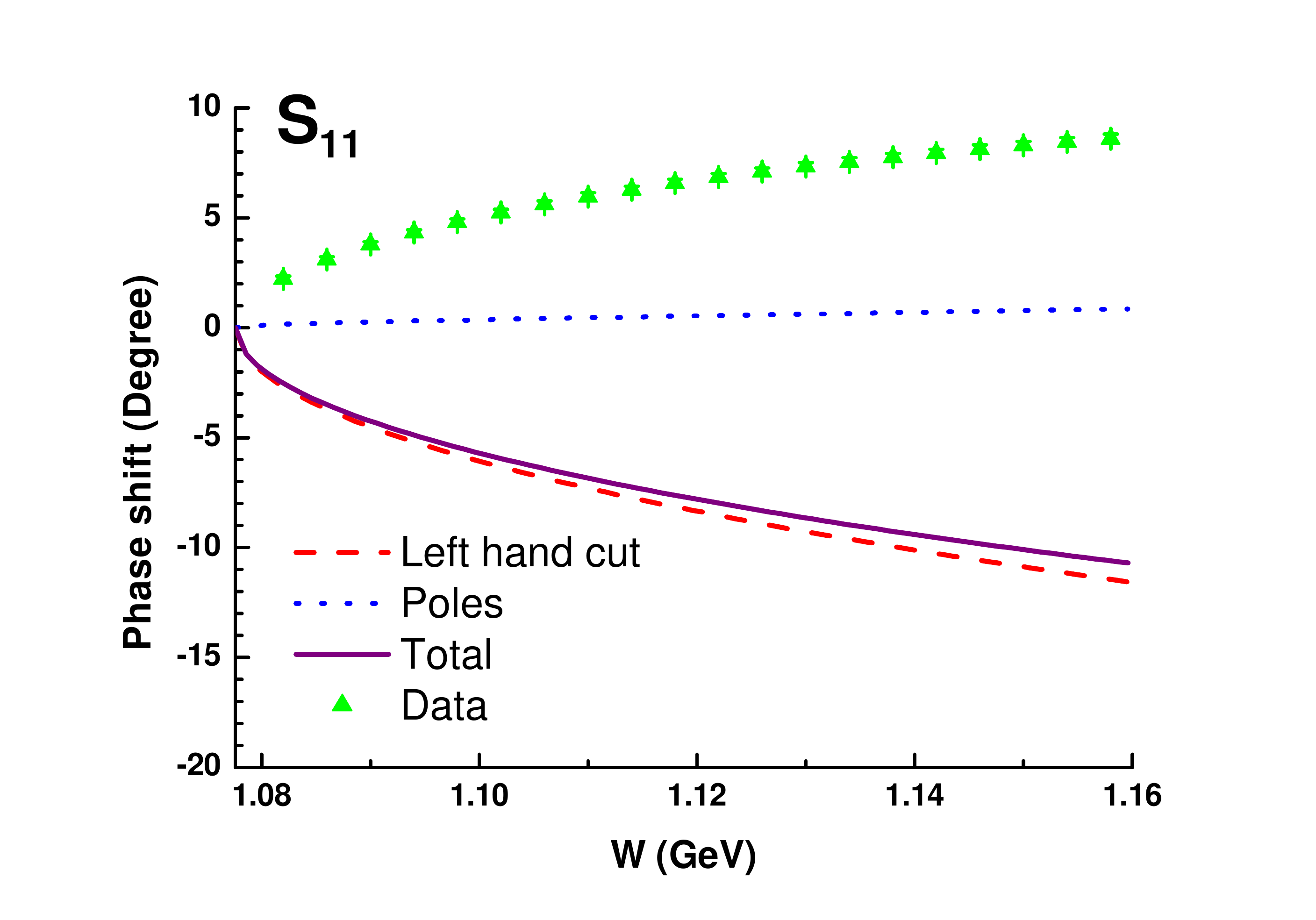}}}
\subfigure[]{
\label{p2PKU:subfig:S31}
\scalebox{1.2}[1.2]{\includegraphics[width=0.4\textwidth]{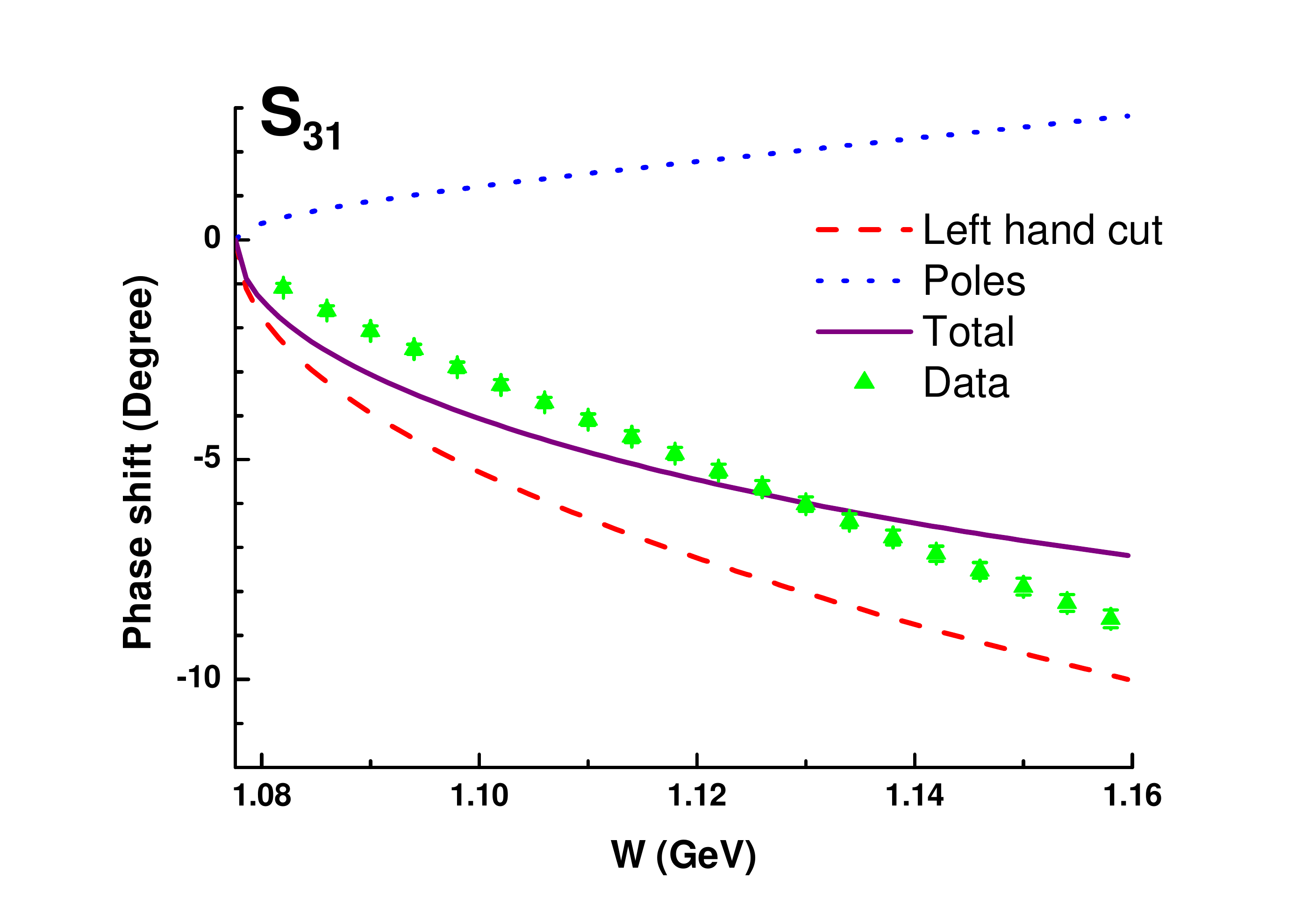}}}
\subfigure[]{
\label{p2PKU:subfig:P11}
\scalebox{1.2}[1.2]{\includegraphics[width=0.4\textwidth]{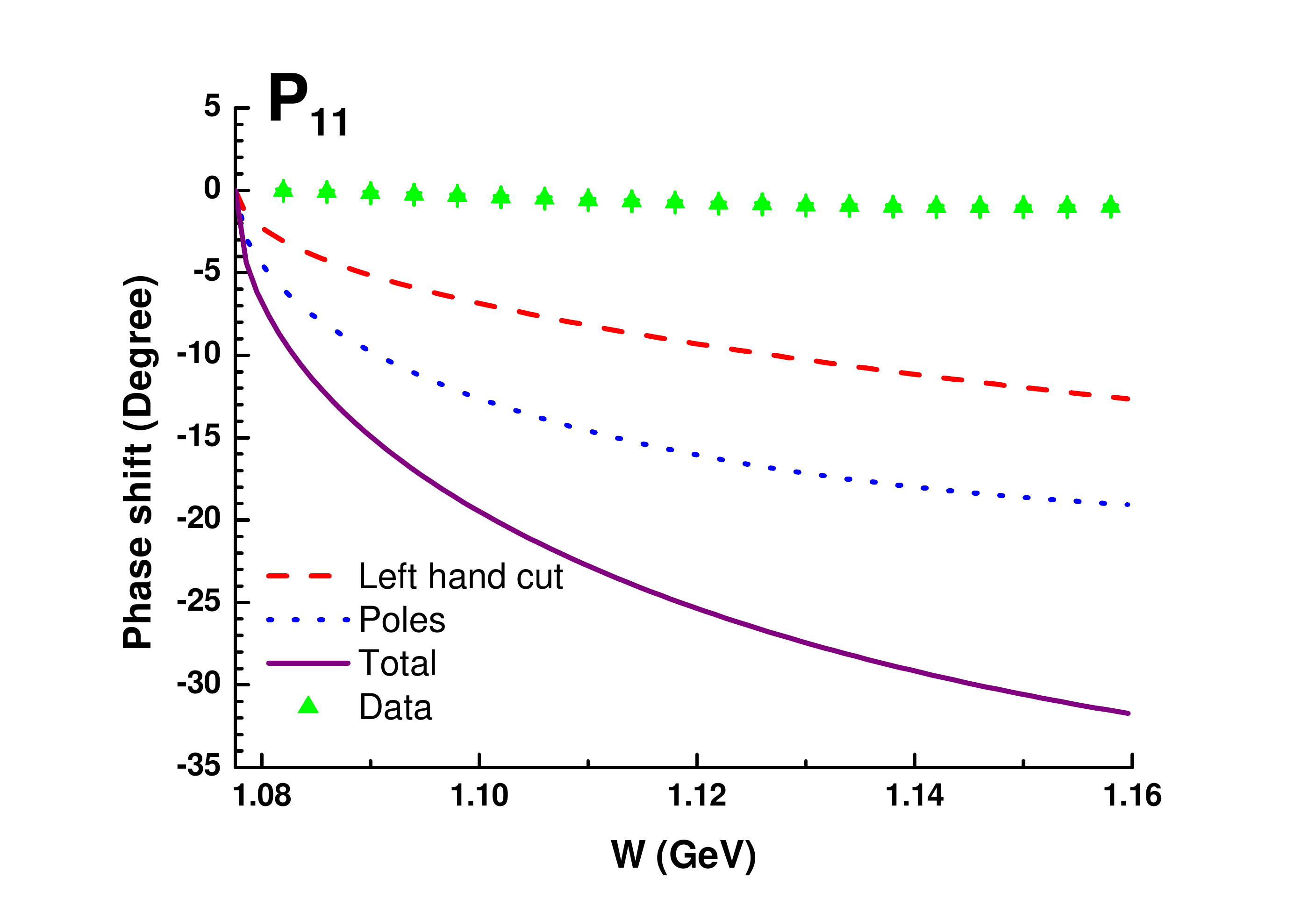}}}
\subfigure[]{
\label{p2PKU:subfig:P31}
\scalebox{1.2}[1.2]{\includegraphics[width=0.4\textwidth]{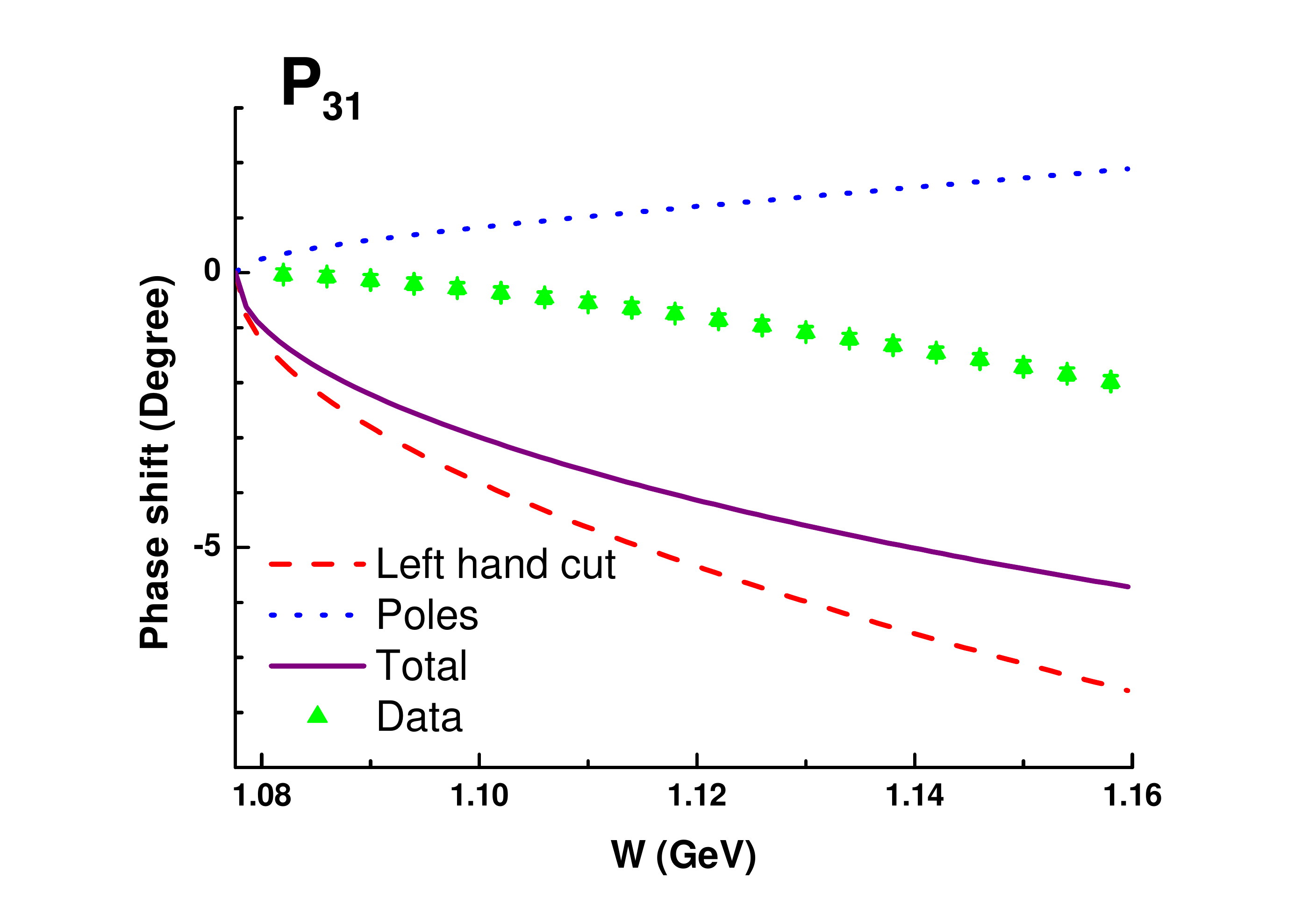}}}
\subfigure[]{
\label{p2PKU:subfig:P13}
\scalebox{1.2}[1.2]{\includegraphics[width=0.4\textwidth]{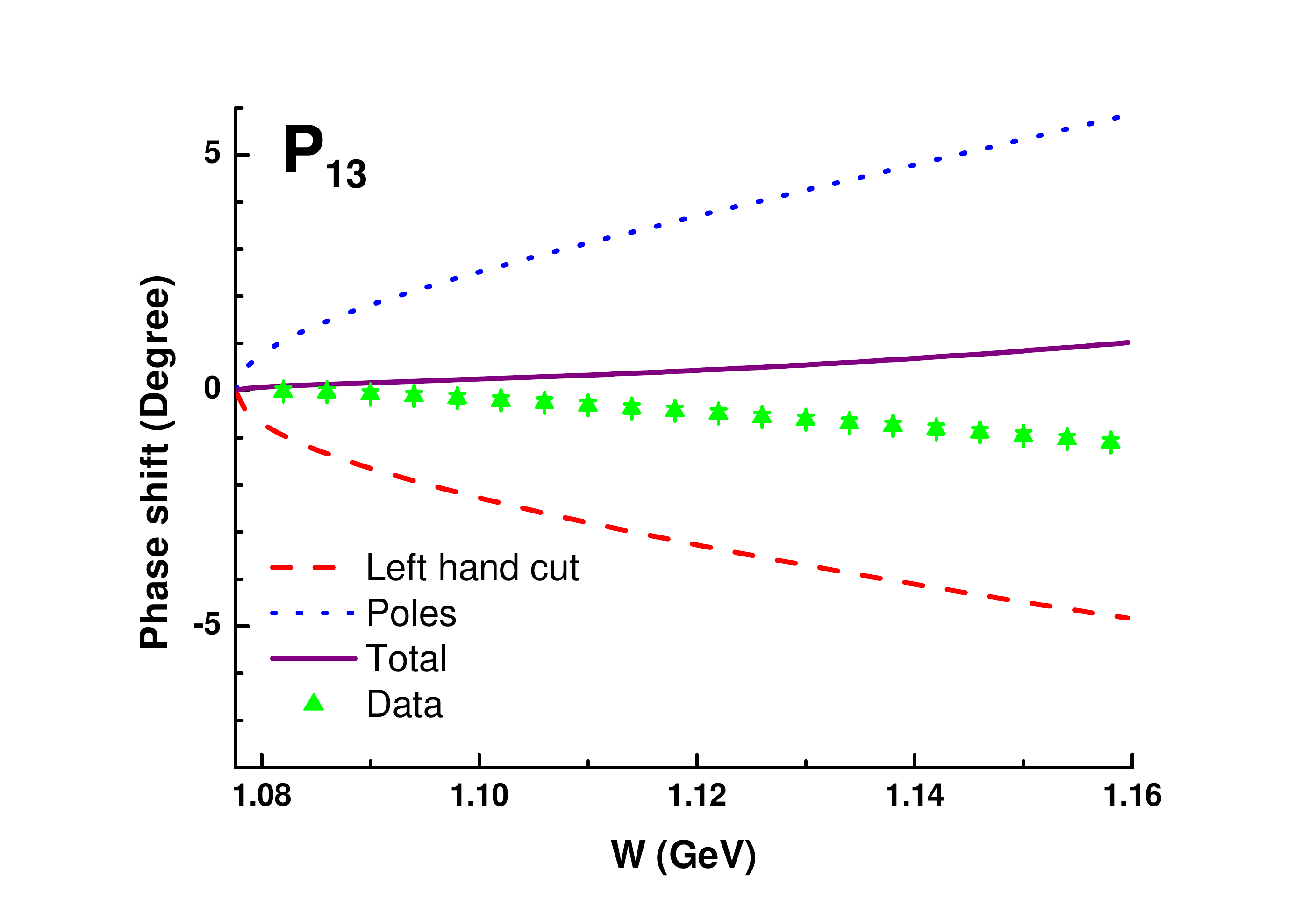}}}
\subfigure[]{
\label{p2PKU:subfig:P33}
\scalebox{1.2}[1.2]{\includegraphics[width=0.4\textwidth]{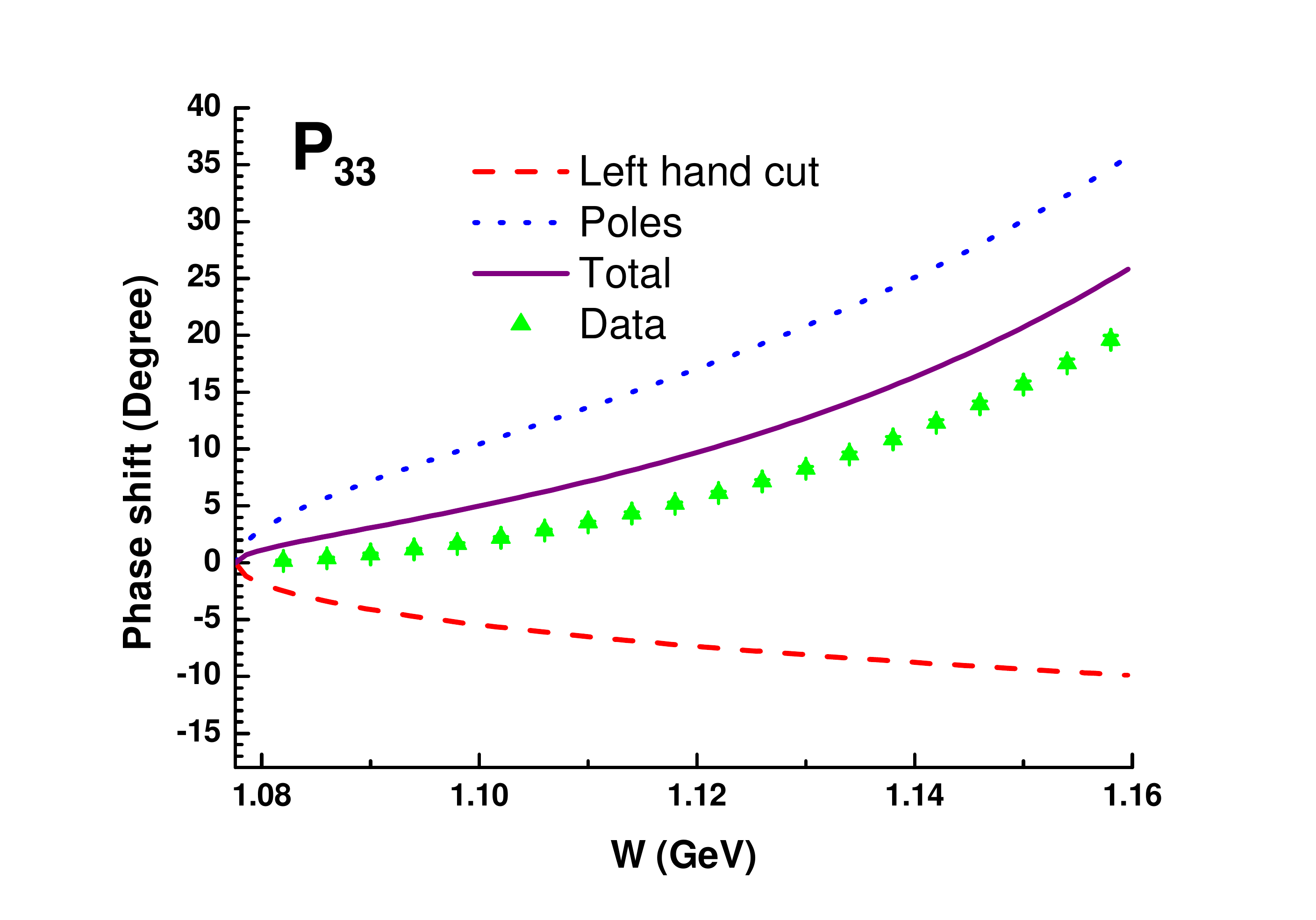}}}
\caption{Tree level PKU representation analyses of the $\pi N$ elastic scattering in $s$ and $p$ waves. }\label{p2PKU}
\end{figure}

The contributions from known poles and \textit{l.h.c.}s to the phase shift in the $S$- and $P$- wave\footnote{In order to separate the contributions clearly, here we do not use free parameters to constraint the near-threshold behavior ($\delta(s)\sim \mathcal{O}(k^3$)) of the $P$- wave amplitudes. As can be seen in Fig.~\ref{p2PKU}, the near-threshold condition can not be satisfied automatically due to the absence of some hidden contributions and the uncertainties in evaluating the \textit{l.h.c.}s. } channels are plotted in Fig.~\ref{p2PKU}, within the scheme of PKU representation.
From Fig.~\ref{p2PKU}, firstly we observe significant contributions from well established poles: in $P_{11}$ channel the nucleon itself serves as a bound state of the $\pi N$ system, generating a large and negative phase shift, while in $P_{33}$ channel the $\Delta(1232)$ resonance gives a large positive contribution. Secondly, the background terms usually contribute sizable and negative phase shifts as expected. Surprisingly, it is found that there exist huge disagreements between the phase shifts from known poles plus cut and the data, especially in $S_{11}$ and $P_{11}$ channels. Therein the results of known poles plus cut have missed some important positive contributions.

One may find that in other channels discrepancies also exist, but are less significant. The discrepancies in $P_{13}$ and $P_{33}$ channels indicate that these channels need a slightly larger $s_c$ parameter; the $S_{31}$ channel may require fine-tuning of $s_c$ parameter and higher order corrections (of the \textit{l.h.c.}); while in $P_{31}$ channel the discrepancy is a bit larger -- an $s_c$ closer to $(M+m)^2$, higher order perturbation calculation or other contributions may be needed. Moreover, changing the $c_i$ parameters may have some impact on those channels. One refers to Appendix.~\ref{app:ci} and~\ref{app:cutoff} for the different choices of $c_i$ and the cut-off parameters. All in all, the discrepancies in those four channels are at quantitative level and may be ascribed to some details of dynamics. On the contrary, the discrepancies in $S_{11}$ and $P_{11}$ channels are qualitatively severe and can never be remedied through such tricks: only when some extra large and positive contributions intrude would they disappear.

There could be some possible interpretations of the huge discrepancies in the $S_{11}$ and $P_{11}$ channels: firstly it is natural to suspect that the one-loop contributions in those two channels may be crucial; secondly, other branch cuts which are not included in the discussions above may be non-ignorable, e.g. the \textit{r.h.i.c.} and the type of cut proposed by Ref.~\cite{Ceci:2011ae}; thirdly, the previous calculation of the shadow poles is carried out under a rough approximation, while the actual shadow pole positions may be very different from what the approximation gives; finally, we can not exclude the existence of some hidden poles, e.g. virtual states, resonances below the threshold, resonances with extremely large widths, or two-pole structures~\cite{Arndt:1985vj}, which cannot be observed directly by the experiments.

As for the first interpretation above, we believe that the higher order contributions of BChPT may have some significance numerically, but they are very unlikely to distort the results totally. As already mentioned before, the background function in Eq.~(\ref{fdisper}) is of logarithmic form and once-subtracted, which would make the function insensitive to the higher order terms. This is also a lesson we learned from $\pi\pi$ scatterings in Ref.~\cite{Zhou:2004ms}. One refers to Appendix.~\ref{app:p1} for the assessment of the dependence of the results on the order of chiral expansion at $\mathcal{O}(p^1)$ and $\mathcal{O}(p^2)$ level. We also extend our calculation of the \textit{l.h.c.} to the $\mathcal{O}(p^3)$ level in $S_{11}$ channel. The result confirms our speculation that the high order corrections to the \textit{l.h.c.} are not qualitatively important. The other interpretations will be discussed in the following two subsections.
\subsection{Estimation of the right-hand inelastic cut}\label{rhccal}
\begin{figure}[htbp]
\center
\subfigure[]{
\label{com:subfig:S11P11rhc}
\scalebox{1.0}[1.0]{\includegraphics[width=0.5\textwidth]{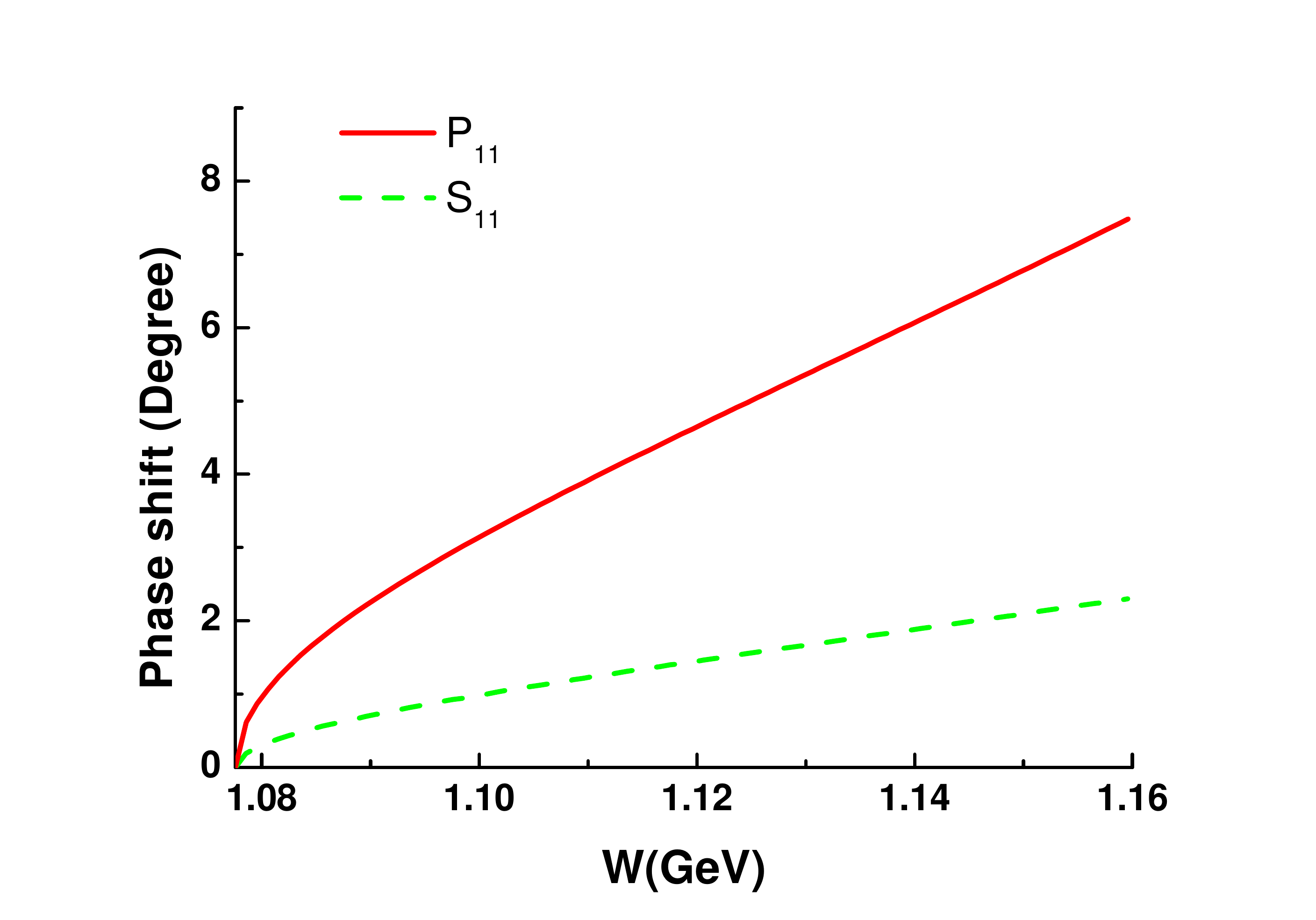}}}
\subfigure[]{
\label{com:subfig:otherrhc}
\scalebox{1.0}[1.0]{\includegraphics[width=0.5\textwidth]{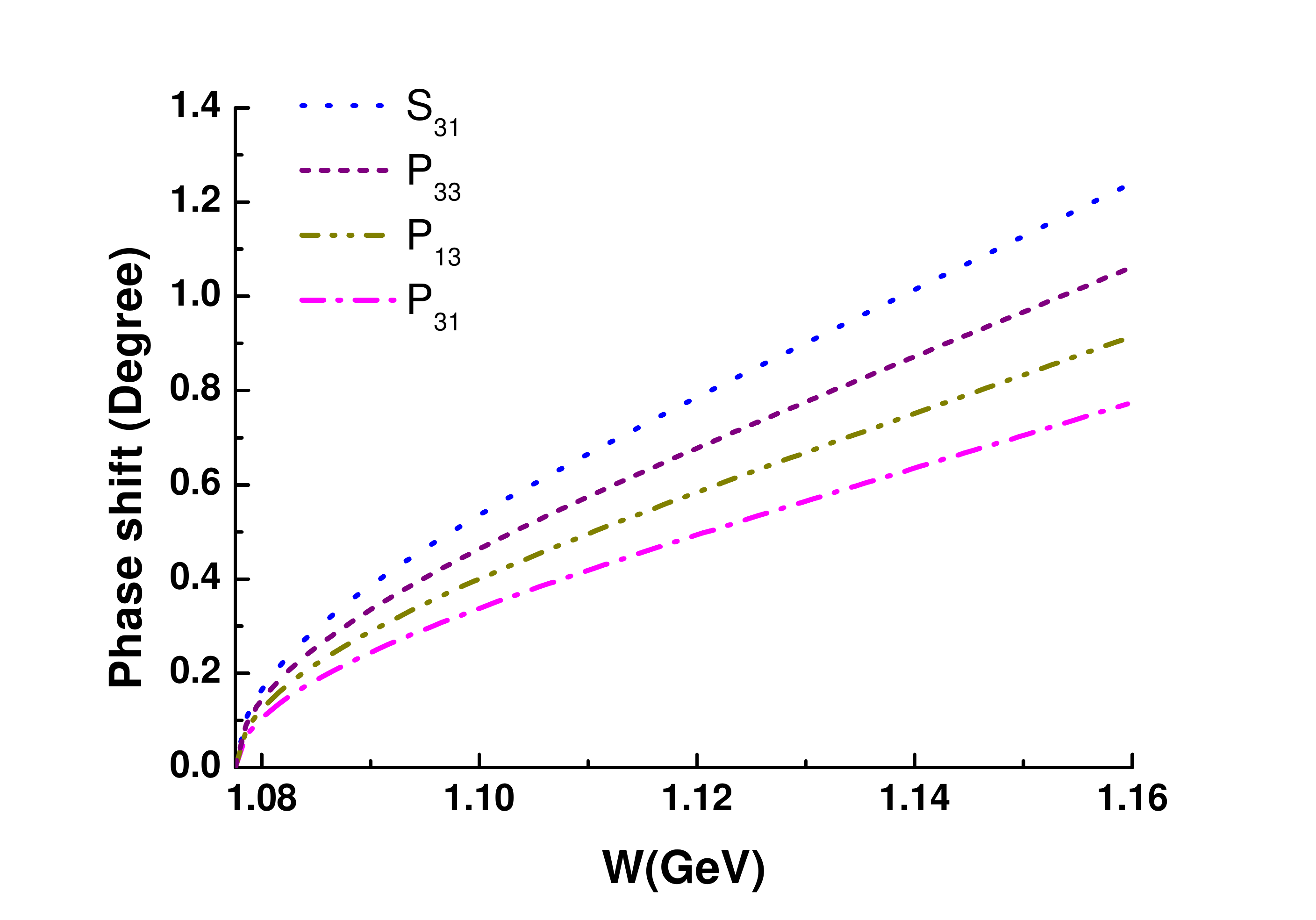}}}
\caption{The phase shifts from \textit{r.h.i.c.}. }\label{fig:rhc}
\end{figure}

The contribution from \textit{r.h.i.c.} is expected to be less significant below the inelastic threshold. Actually, analogous to Eq.~\eqref{fsint}, the contribution from the \textit{r.h.i.c.} can be estimated by
\begin{equation}\label{fsRint}
\begin{split}
&f_{\text{R}'}(s)=\frac{s}{\pi}\int_{(2m+M)^2}^{\Lambda_R^2} \frac{\sigma_R(w)dw}{w(w-s)}\ \mbox{, }\\
&\sigma_R(w)=-\Big\{\frac{\ln[\eta(w)]}{2\rho(w)}\Big\}\ \mbox{, }
\end{split}
\end{equation}
where $0\leq\eta\leq 1$ represents the inelasticity of $\pi N$ scatterings above the inelastic threshold, i.e. the $\pi\pi N$ threshold $(2m+M)^2$, and the cut-off parameter of the integral in Eq.~\eqref{fsRint} is denoted as $\Lambda_R$. Eq.~\eqref{fsRint} indicates that the \textit{r.h.i.c.} contribution is positive definite. In the following computation, the inelasticity function $\eta(s)$ is {taken from} Ref.~\cite{SAID}. As for the cut-off $\Lambda_R$, the energy region of the data is only up to $2.48$ GeV, but in $P_{11}$ channel the inelasticity is still near $0.1$ when the centre of mass energy reaches $2.48$ GeV. Hence to avoid the disappreciation of right-hand cut contribution, we take a cut-off  of $\Lambda_R=4.00$ GeV and employ an extrapolation of the data to $4.00$ GeV to estimate the function $f_{\text{R}'}$ in Eq.~\eqref{fsRint}. The phase shifts from inelastic cuts of different channels are plotted in Fig.~\ref{fig:rhc}.
It is clear to see that in $P_{11}$ channel the \textit{r.h.i.c.} contributes sizeably to the phase shift (about $8$ degrees at $\sqrt{s}=1.16$ GeV), but this is still far from enough to compensate the discrepancy (nearly $32$ degrees at $\sqrt{s}=1.16$ GeV); meanwhile, the phase shifts from \textit{r.h.i.c.} in other channels are much smaller, thus the discrepancy of $S_{11}$ channel can neither be interpreted by it.

Besides, to investigate other possible hidden cut structures (as suggested by Ref.~\cite{Ceci:2011ae}) goes beyond the scope of present paper.
\subsection{Hidden pole analyses: $P_{11}$ channel}
After all the discussions above, it seems that the existence of hidden poles is the last remaining interpretation. It is noticed that we can not tell the differences between shadow pole positions very different from Eq.~(\ref{shadowpole}) and extra hidden poles, because shadow pole positions determined by Eq.~(\ref{shadowpole}) contributes little to the phase shifts at low energies\footnote{This argument also indicates that the two-pole structures under Eq.~\ref{shadowpole} cannot explain the discrepancy. }; however, the missing positive contributions in $S_{11}$ and $P_{11}$ channels are too large to be filled by shadow pole positions close to what Eq.~(\ref{shadowpole}) gives, so it is more natural to believe that the discrepancy stems from extra hidden poles, which are discussed as follows.

\begin{figure}[htbp]
\centering
\includegraphics[width=0.6\textwidth]{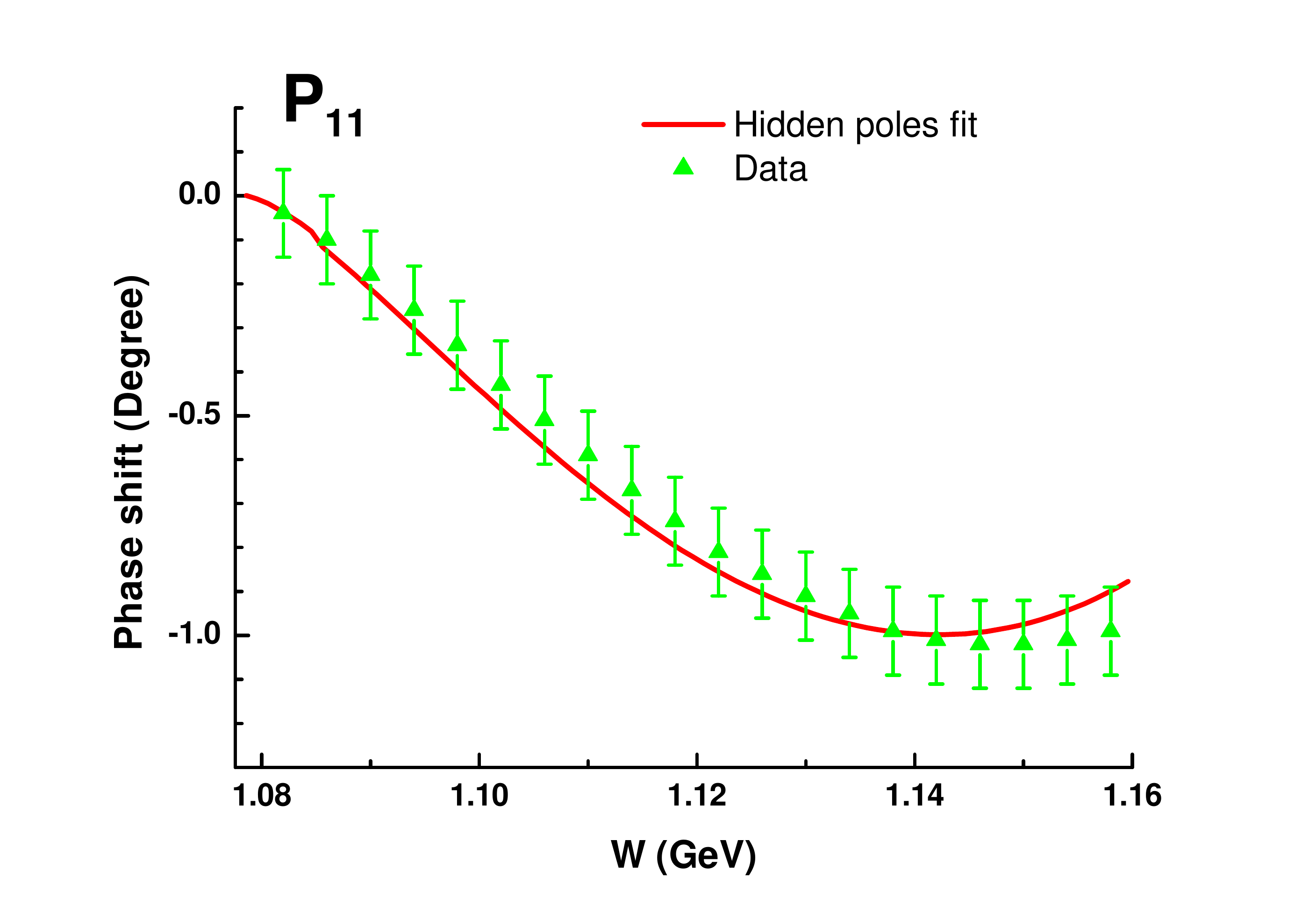}
\caption{The phase shift results with extra virtual states in $P_{11}$ channel. }
\label{fig:P11p2sh}
\end{figure}

We proceed by keeping the known poles and the inelastic cut, meanwhile we still keep the values of $c_i$ coefficients in Eq.~\eqref{civalue}. Threshold $P$- wave behavior is used as a constraint: the phase shift cannot contain $\mathcal{O}(k^1)$ term (here $k$ is the $3$-momentum of the $\pi N$ system in center of mass frame). It is found that if we set the extra pole to be one resonance and perform a fit to the data, it always automatically runs to the real axis and become two virtual poles: one survives while the other falls to the pseudo-threshold ($\sqrt{s}=M-m$) and vanishes. The $P_{11}$ data requires passionately one single virtual state near the $\pi N$ threshold. Notice that both the resonance pole and the virtual pole give similar positive phase shift, it is amazing that the fit using PKU representation can easily distinguish the two very similar contributions. This is very remarkable -- as will be revealed later, that a single nearby virtual state is actually what is needed physically. It is found that in $P_{11}$ channel a good fit requires a much larger $s_c$ than $-0.08$ GeV$^2$, but a nearby virtual state always exists irrespective of the choice of the cut-off parameter\footnote{When $s_c=-0.08$ GeV$^2$, the fit quality is poor, but the $P$-wave constraint gives a virtual state at $952$ MeV, which is above the nucleon pole and not far from the location when $s_c=-9$ GeV$^2$, which generates Fig.~\ref{fig:P11p2sh}. }. Fig.~\ref{fig:P11p2sh} affords an example with $\chi^2_{P_{11}}/\text{d.o.f}=0.240$, where the virtual pole locates at $983$ MeV.

In the following we show that through investigating the amplitude, the existence of that pole is found to be quite reasonable -- actually, it is just a kinematic companionate pole of the nucleon bound state. In the vicinity of the nucleon bound state pole, the $S$ matrix takes the form
$$S(s)\sim \frac{r_0}{s-M^2}+b_0+\mathcal{O}(s-M^2)$$
where $r_0\in\mathbb{R}$ is the residue and the constant $b_0$ represents the background. The dominant term ${r_0}/{(s-M^2)}$ is a typical hyperbola with the horizontal axis being one of its asymptotes, while arbitrary non-zero real $b_0$ would lead the function to generate a zero on the first Riemann sheet, which, according to the rule of analytic continuation $S^{\text{II}}=1/S^{\text{I}}$, indicates the existence of a virtual state pole on the second sheet.\footnote{Exactly speaking, $b_0$ is complex due to the existence of the $u$-channel nucleon exchange. Nevertheless, since the $u$-channel cut is numerically very small, the main conclusion of the observation here remains unchanged. } Such a simple explanation on the necessity of a nearby virtual state, is firstly revealed here, to the best of our knowledge. For example, in Ref.~\cite{Lutz:2010}, a somewhat arbitrary CDD pole is introduced to take the role of such a virtual state without further investigation of its origin.

\begin{figure}[htbp]
\centering
\includegraphics[width=0.6\textwidth]{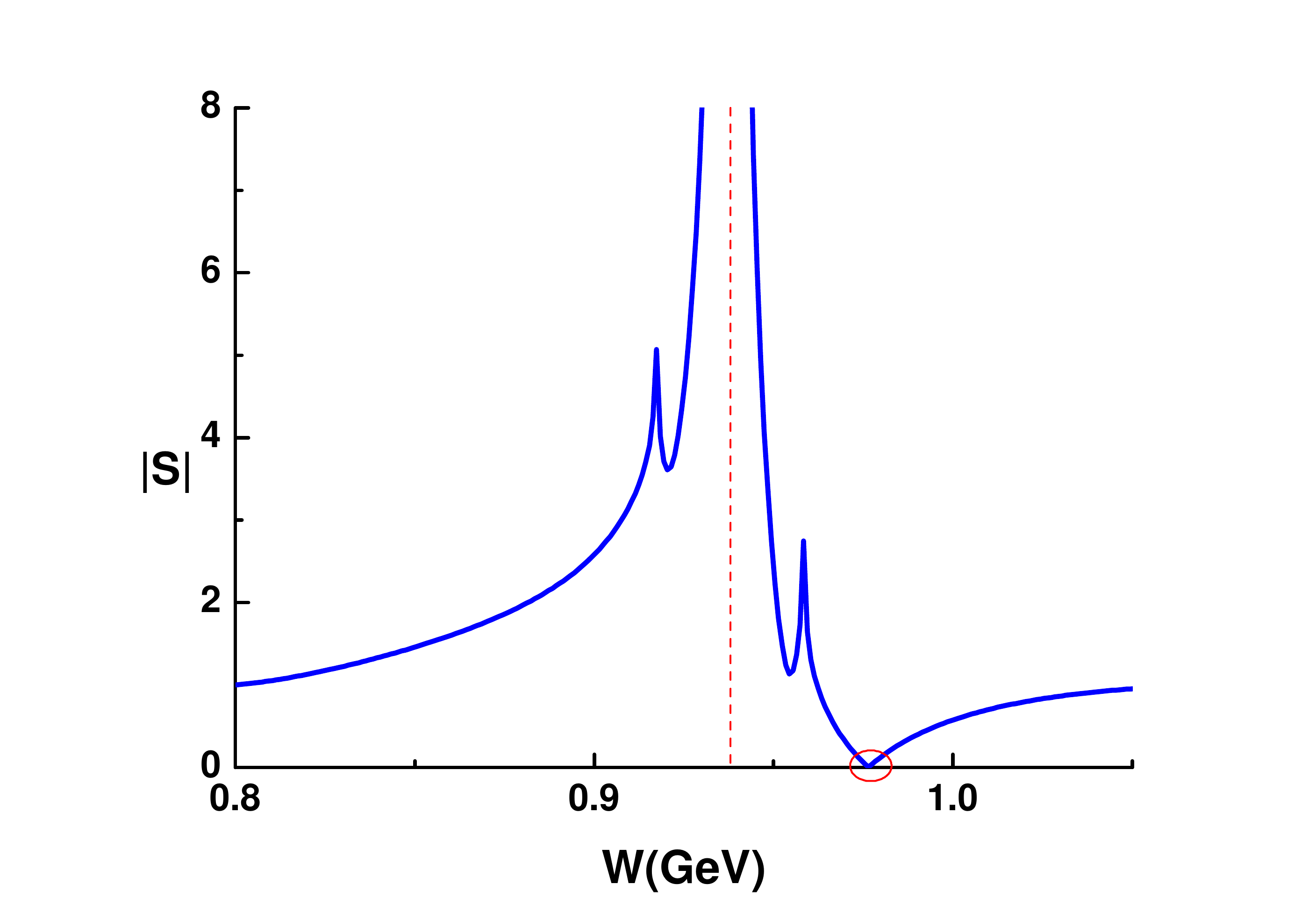}
\caption{The absolute value of perturbative calculated $S$ matrix in $P_{11}$ channel below threshold. The vertical red dashed line marks the position of the nucleon bound state pole, and the red circle illustrates the zero of the $S$ matrix. The two spikes correspond to the two branch points of the $u$-channel cut. }
\label{fig:P11perS}
\end{figure}

On the other hand, the partial-wave $S$ matrix given by perturbative calculation also suggests a zero at
\begin{equation}
\left[M^2+\frac{2g^2m^3M}{3F^2\pi}-\frac{g^2m^6(3+4g^2-4c_3 M+8c_4 M)}{18F^4\pi^2}+\cdots\right]^{1/2}\sim 976\ \text{MeV, }
\end{equation}
see Fig.~\ref{fig:P11perS}, which is, as it should be, close to the result $983$ MeV given by the fit of PKU representation with the $P$- wave constraint in Fig.~\ref{fig:P11p2sh}.

Remember that the virtual pole resides in the domain of validity of chiral expansions, the above analyses demonstrates that the PKU representation is very sensitive to the low lying pole positions and hence is very powerful and reliable in pinning down the correct low-energy pole positions, which encourages us to find out the hidden contribution in $S_{11}$ channel similarly.
\subsection{Hidden pole analyses: $S_{11}$ channel}
For $S_{11}$ channel, we keep Eq.~(\ref{shadowpole}) as a solution for $N^*(1535)$ shadow pole position, and add one more resonance with mass and width as free parameters, meanwhile we keep $s_c=-0.08$ GeV$^2$. Unlike $P_{11}$ channel, our fit leads to a resonance pole lying below the $\pi N$ threshold:{$\sqrt{s}^{\text{II}}=0.808-0.055i$ GeV, with the fit quality $\chi_{S_{11}}^2/\text{d.o.f}=0.109$.} Notice that this pole location may not be that accurate even though we have no little doubt on the very existence of such a pole. The dependence of the pole location on the variation of background contribution is analyzed in Appendix.~\ref{app:cutoff}. For example, taking $s_{c}=-\infty$, one gets $\sqrt{s}^{\text{II}}=0.914-0.205i$ GeV, which still locates well below threshold, with a similar fit quality $\chi^2_{S_{11}}/\text{d.o.f}=0.018$. From these analyses, we see that the vast change of cut-off parameter does not lead to a significant change of the pole position. From Table.~\ref{tab:S11sh} we estimate the pole location to be
\begin{equation}\label{S11shp}
\sqrt{s}^{\text{II}}=(0.861\pm0.053)-(0.130\pm0.075)i\ \mbox{GeV. }
\end{equation}
\begin{figure}[htbp]
\centering
\includegraphics[width=0.6\textwidth]{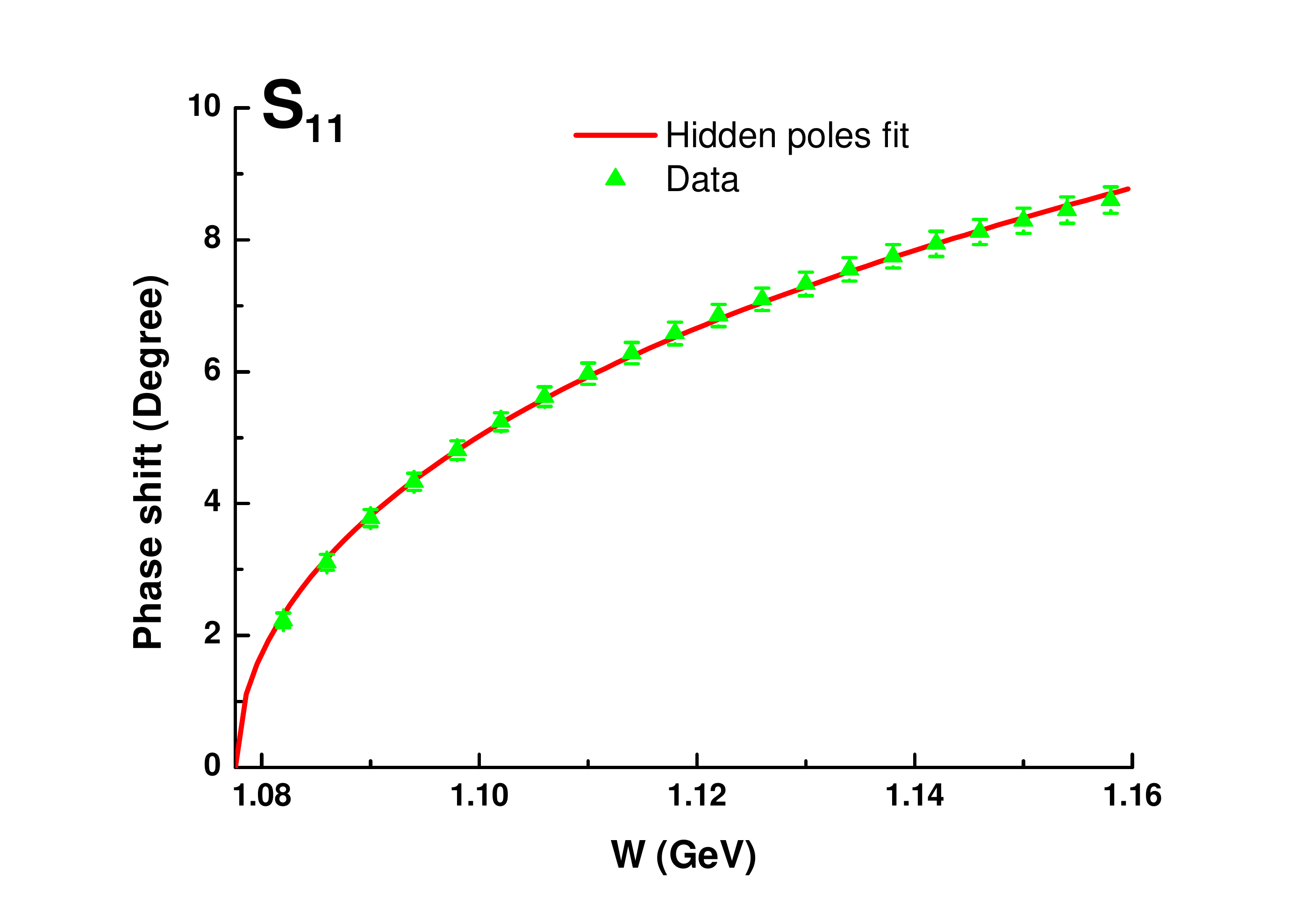}
\caption{The hidden pole fit to data in $S_{11}$ channel. The background contribution can be read off from Fig.~\ref{p2PKU:subfig:S11}. The crazy resonance saturates the gap between the data and the background contribution. }
\label{fig:S11p2sh}
\end{figure}

\begin{figure}[htbp]
\centering
\includegraphics[width=0.6\textwidth]{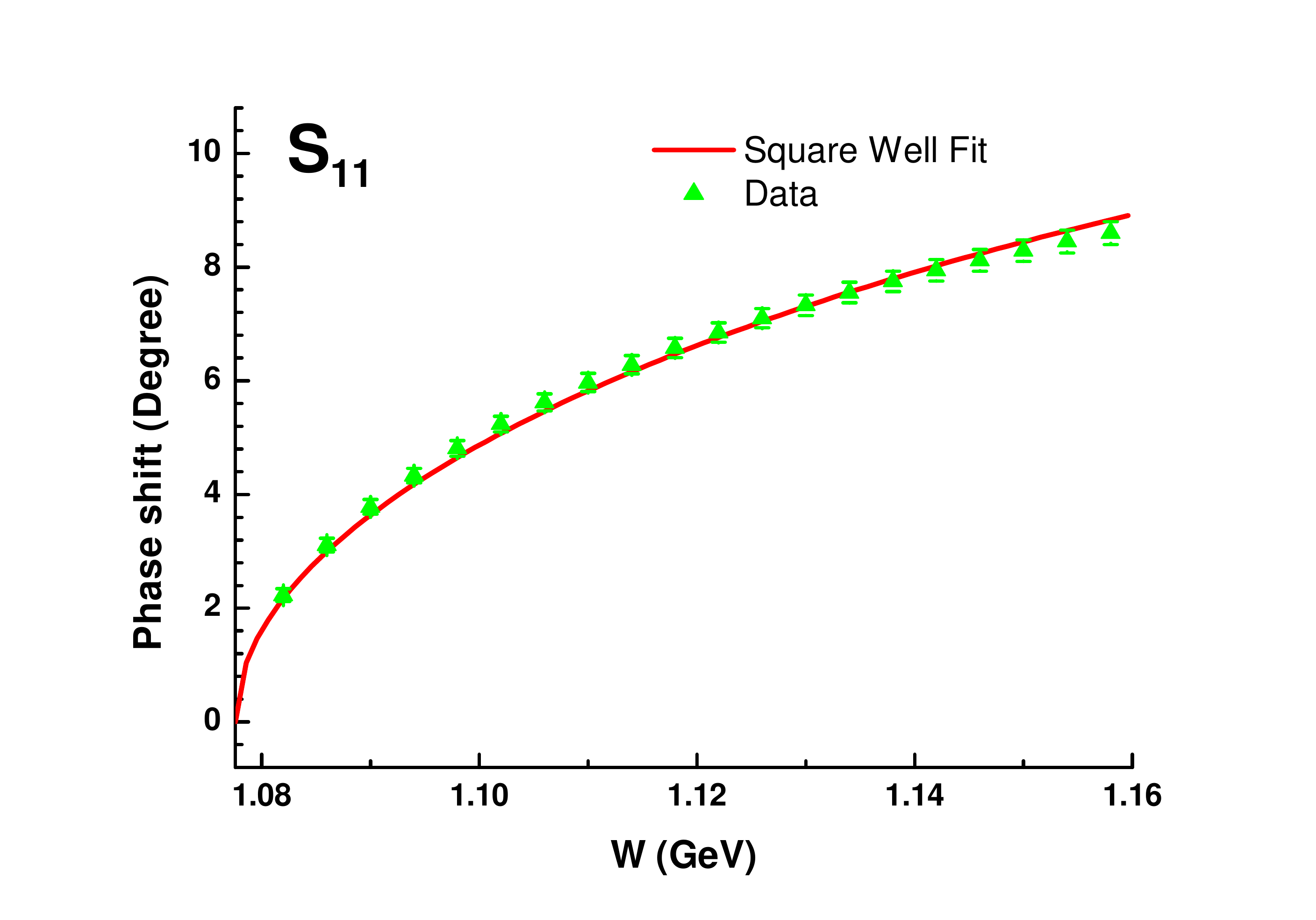}
\caption{The fit to the data with phase shift given by square well potential scatterings. }
\label{fig:S11sw}
\end{figure}

\begin{figure}[htbp]
\centering
\includegraphics[width=0.6\textwidth]{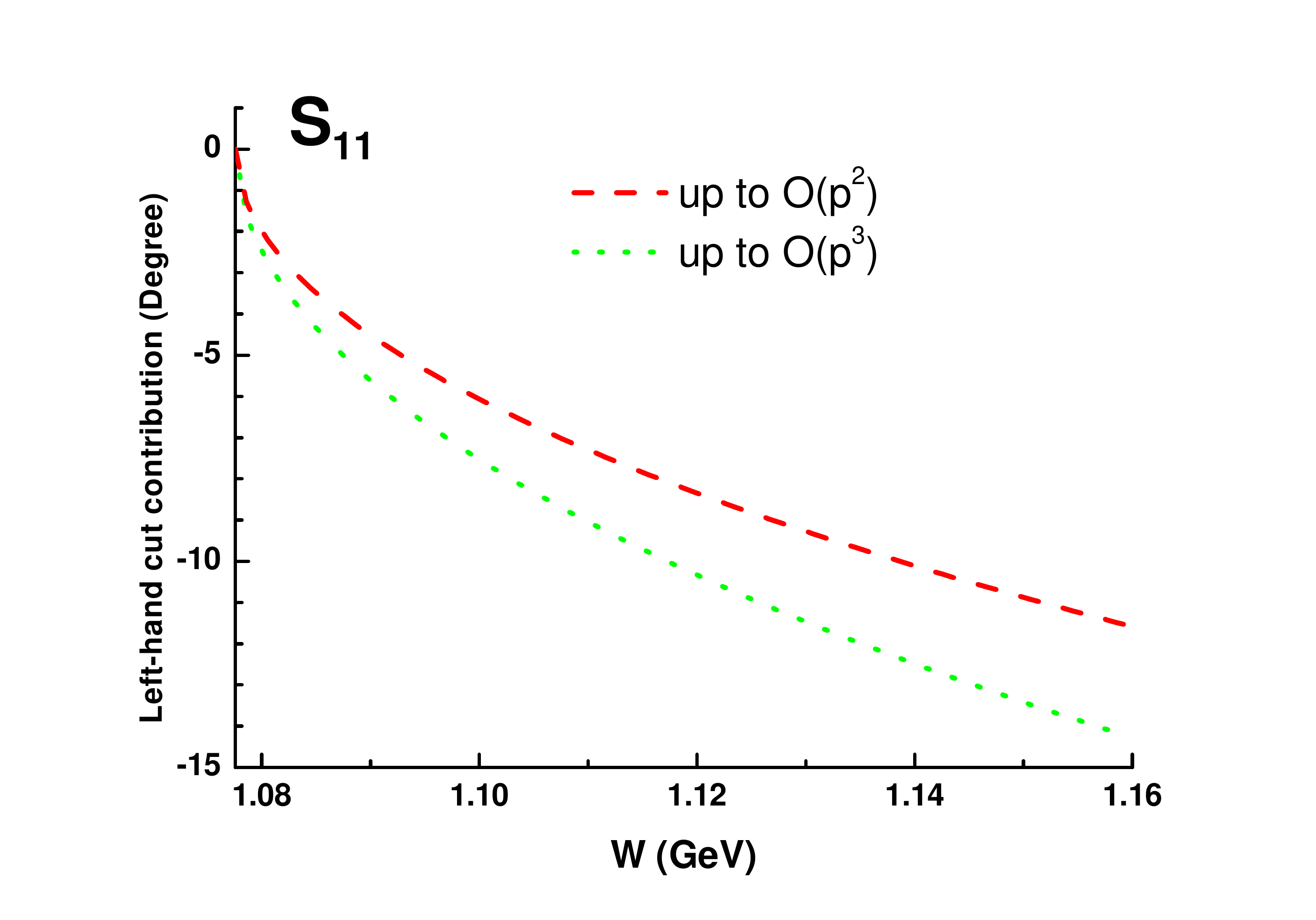}
\caption{The left-hand cut contributions in $S_{11}$ channel up to $\mathcal{O}(p^2)$ and $\mathcal{O}(p^3)$. The cut-off parameters are determined by the $N^*(1440)$ shadow pole location, and the values of the low energy constants are taken from Table. 1 (Fit II) in Ref.~\cite{Chen:2012nx}. }
\label{fig:S11p23}
\end{figure}

The hidden resonance found above is beyond direct experimental observations and is difficult to be understood. However, we would try to give it a reasonable interpretation based on our past experiences about potential scatterings. The $S_{11}$ channel does not contain the nucleon as an $s$ channel intermediate bound state, so the interaction is dominated by contact interaction and $u$ channel nucleon exchange, which can be simulated as a potential. Typical potential scatterings can generate resonances below threshold; for example, we use the square-well potential as a toy model to show how this happens. The potential is
\begin{equation}
U(r)=2\mu V(r)=
\begin{cases}
-2\mu V_0 &\mbox{($r\leq L$), }\\
0 &\mbox{($r>L$), }
\end{cases}
\end{equation}
where $\mu$ is the reduced mass of the pion and nucleon, $r$ is the radial coordinate, $V_0>0$ and $L>0$ are two parameters labeling the depth and range of the potential respectively. The textbook calculation of $S$- wave Schr\"{o}dinger equation gives the phase shift, as a function of $3-$momentum $k$ in centre of mass frame:
\begin{equation}
\delta_{\text{sw}}(k)=\arctan\left[\frac{k\tan{k'L}-k'\tan{kL}}{k'+k\tan{(kL)}\tan{(k'L)}}\right]\ \mbox{, }
\end{equation}
with $k'=(k^2+2\mu V_0)^{1/2}$. The fit to the data in $S_{11}$ channel (20 data points) results in $L=0.829$ fm and $V_0=144$ MeV, with fit quality $\chi^2_{\text{sw}}/\text{d.o.f}=0.740$, see Fig.~\ref{fig:S11sw}.
The poles are extracted via the equation $1-i\tan\delta_{\text{sw}}(k)=0$, the nearest one of which locates at $k=-346i$ MeV as a deep virtual state, which in $\sqrt{s}$ plane turns out to be a crazy resonance\footnote{The second pole given by square-well potential is at $2.28-1.06i$ GeV, which is too far away from the expected $N^*(1535)$ position.}:
\begin{equation}
\sqrt{s}=\sqrt{k^2+m^2}+\sqrt{k^2+M^2}\sim 0.872-0.316i\ \mbox{GeV. }
\end{equation}
Note that the deep virtual state suggested by the square well potential is already out of the non-relativistic energy region, but it still agrees with the result of PKU representation analyses (Eq.~\ref{S11shp}) qualitatively. Even though the square well potential calculation is only a toy model analysis, one expects it reveals the correct physical picture that the crazy resonance is of potential scattering nature, generated from a not very strong but attractive potential. However it should be warned that the hidden pole position in $S_{11}$ channel is inside the circular left-hand cut $\text{Re}(s)^2+\text{Im}(s)^2=(M^2-m^2)^2$, which, though only appears when the one loop contributions are under consideration, indicates that the one loop results may have critical impact on the hidden pole. One may even doubt that such a hidden pole is only a fake effect simulating the circular cut contribution, which are not presented at the moment. Nevertheless, as already discussed in Section.~\ref{sec:PKU}, higher order calculations to the left-hand cut should not alter the qualitative picture presented here. This expectation is supported by the crucial calculation at $\mathcal{O}(p^3)$ level~\cite{Wang:2018}. Briefly speaking, our $\mathcal{O}(p^3)$ preliminary result shows the hidden pole remains below threshold, since the left-hand cut contribution up to $\mathcal{O}(p^3)$ in $S_{11}$ channel differs little from $\mathcal{O}(p^2)$, see Fig.~\ref{fig:S11p23}.
\section{Conclusions and outlook}\label{con}
In this work, we apply PKU representation, which separates the phase shifts into terms corresponding to different poles and branch cuts, to analyse processes of $\pi N$ elastic scatterings. The contribution of background term, i.e. left-hand cut contribution, is deduced from tree amplitude derived in manifestly covariant BChPT. It is found that the left-hand cut in each $S$- or $P$- wave channel contributes negatively to the phase shift, in agreement with conventional wisdom.

Having estimated the left and right-hand cut contributions, it is found, particularly in $P_{11}$ and $S_{11}$ channels, the total contribution of the known poles and branch cuts are apparently insufficient to yield a satisfactory description of the experimental data. Thus certain significant positive contributions are required to compensate the discrepancies in those two channels.

In $P_{11}$ wave, with the assistance of PKU representation, a kinematical near-threshold virtual state induced by the nucleon bound state pole is discovered, the location of which is found to be compatible with that calculated from perturbation theory. The discovery of that pole demonstrates the uncertainties from left-hand cut contributions extracted from the perturbation amplitudes is well under control in the framework of PKU representation. Further, the origin of such a virtual pole has not been discussed in the literature before.

Assuming that the $S_{11}$ channel includes an extra hidden pole, we determine its position on the second Riemann sheet by fitting to phase shift data. It is found that the $S_{11}$ channel may cache a pole lying well below the $\pi N$ threshold, behaving as a crazy resonance. To reveal the existence of such a pole is completely novel.

There are still many follow-ups of this work to be done in future. For instance, the calculation of the left-hand cut integral could be extended to $\mathcal{O}(p^3)$ level, which is underway; methods based on crossing symmetry, e.g. Roskies relation or Roy-Steiner equations~\cite{Ditsche:2012fv}, may be incorporated into to get a cross-check on the determination of pole location in $S_{11}$ channel. The physics related to such a novel resonance remains to be explored.
\\
\\
{\it Acknowledgments:} This work is supported in part by National Nature Science Foundations of China (NSFC) under Contract Nos. 10925522, 11021092, by the Spanish Ministerio de Econom\'ia y Competitividad and the European Regional Development Fund, under contracts FIS2014-51948-C2-1-P, FIS2014-51948-C2-2-P, SEV-2014-0398 and by Generalitat Valenciana under contract PROMETEOII/2014/0068. DLY acknowledges the hospitality of the ITP of CAS where part of this work was done.

\appendix{\center\bf\huge Appendices}
\section{Tree-level $A,B$ functions}\label{app:ff}
The expressions of the $A,B$ functions can be found elsewhere, e.g., Refs.~\cite{Alarcon:2012kn,Chen:2012nx}. For completeness, we show the results as follows. At $\mathcal{O}(p^1)$ level
\begin{align}
&A_1^{1/2}=\frac{g^2M}{F^2}\ \mbox{, }\\
&B_1^{1/2}=\frac{1-g^2}{2F^2}-\frac{3M^2g^2}{F^2(s-M^2)}-\frac{M^2g^2}{F^2}\frac{1}{u-M^2}\ \mbox{, }\\
&A_1^{3/2}=\frac{g^2M}{F^2}\ \mbox{, }\\
&B_1^{3/2}=-\frac{1-g^2}{2F^2}+\frac{2M^2g^2}{F^2(s-M^2)}\ \mbox{; }
\end{align}
and for $\mathcal{O}(p^2)$
\begin{align}
&A_2^{1/2}=-\frac{4c_1 m^2}{F^2}+\frac{c_2(s-u)^2}{8M^2F^2}+\frac{c_3}{F^2}(2m^2-t)-\frac{c_4(s-u)}{F^2}\ \mbox{, }\\
&B_2^{1/2}=\frac{4Mc_4}{F^2}\ \mbox{, }\\
&A_2^{3/2}=-\frac{4c_1 m^2}{F^2}+\frac{c_2(s-u)^2}{8M^2F^2}+\frac{c_3}{F^2}(2m^2-t)+\frac{c_4(s-u)}{2F^2}\ \mbox{, }\\
&B_2^{3/2}=-\frac{2Mc_4}{F^2}\ \mbox{, }
\end{align}
where the subscripts denote the chiral orders.
\section{Partial wave helicity amplitudes}\label{app:pw}
In this section, we express the explicit expressions of $\pi N$ partial wave amplitudes at tree level, which presents the left-hand cut structure clearly. To the best of our knowledge, these expressions have never been exhibited in previous literature. In what follows, we use the abbreviations: $R_m=M^2-m^2$, $R_p=M^2+m^2$, $c_L=(M^2-m^2)^2/M^2$ and $c_R=M^2+2m^2$. The kinematic factor $\rho(s)$ is given by Eq.~(\ref{rhodef}).
\subsection{$\mathcal{O}(p^1)$ amplitudes}
The $\mathcal{O}(p^1)$ partial wave helicity amplitudes are written as
\begin{align}
&T_{++}^{I,J}=\mathfrak{A}^{I,J}+\mathfrak{B}^{I,J}I_C^J(s)\ \mbox{, }\label{TppAB}\\
&T_{+-}^{I,J}=\frac{1}{\sqrt{s}}(\mathfrak{C}^{I,J}+\mathfrak{D}^{I,J}I_S^J(s))\ \mbox{, }\label{TpmCD}
\end{align}
with definite isospin $I$ and angular momentum $J$. In Eqs.~(\ref{TppAB}) and (\ref{TpmCD}), the $I_{C,S}$ functions corresponding to the $u$-channel nucleon exchange are singled out from the partial wave helicity amplitudes with the coefficients $\mathfrak{A}$, $\mathfrak{B}$, $\mathfrak{C}$ and $\mathfrak{D}$. The $I_{C,S}$ functions have the form of
\begin{align}
&I_C^{1/2}(s)=\int_{-1}^1\frac{1+z_s}{2(u-M^2)}dz_s\ \mbox{ }\nonumber\\
&=-\frac{2}{s^2\rho^4}\Big[\frac{M^2}{s}(s-c_L)\Big(\ln\frac{M^2}{s}+\ln\frac{s-c_L}{s-c_R}\Big)+s\rho^2\Big]\ \mbox{, }\\
&I_S^{1/2}(s)=\int_{-1}^1\frac{1-z_s}{2(u-M^2)}dz_s\ \mbox{ }\nonumber\\
&=\frac{2}{s^2\rho^4}\Big[(s-c_R)\Big(\ln\frac{M^2}{s}+\ln\frac{s-c_L}{s-c_R}\Big)+s\rho^2\Big]\ \mbox{, }
\end{align}
for $J=1/2$ and
\begin{align}
&I_C^{3/2}(s)=\int_{-1}^1\frac{(1+z_s)(3z_s-1)}{4(u-M^2)}dz_s\ \mbox{ }\nonumber\\
&=\frac{1}{s^5\rho^6}\Big\{2M^2(s-c_L)\big[2s(s-m^2)-M^2c_L-sc_R\big]\Big(\ln\frac{M^2}{s}+\ln\frac{s-c_L}{s-c_R}\Big)\nonumber\\
&+s^2\rho^2(s^2-2m^2s-5M^2c_L+4M^2s)\Big\}\ \mbox{, }\\
&I_S^{3/2}(s)=\int_{-1}^1\frac{(1-z_s)(3z_s+1)}{4(u-M^2)}dz_s\ \mbox{ }\nonumber\\
&=-\frac{1}{s^4\rho^6}\Big\{2(s-c_R)(s^2+M^2s-2m^2s-2M^2c_L)\Big(\ln\frac{M^2}{s}+\ln\frac{s-c_L}{s-c_R}\Big)\ \mbox{ }\nonumber\\
&+s\rho^2\big[4(s-c_R)s-2sm^2+s^2-M^2c_L\big]\Big\}\ \mbox{, }
\end{align}
for $J=3/2$, where the Mandelstam variable $u$ is given by
\begin{equation}
u(s,z_s)=R_p-\frac{s^2-R_m^2}{2s}-\frac{(s-s_L)(s-s_R)}{2s}z_s\ \mbox{. }
\end{equation}
The expressions of $\mathfrak{A}, \mathfrak{B}, \mathfrak{C}$ and $\mathfrak{D}$ in Eqs.~(\ref{TppAB}) and (\ref{TpmCD}) are as follows.
\begin{itemize}
\item{$I=1/2, J=1/2$}
\begin{align}
&\mathfrak{A}^{1/2,1/2}=\frac{s-R_p}{32F^2\pi}-\frac{g^2(s^2-R_ps-2m^2M^2)}{32F^2\pi(s-M^2)}\ \mbox{, }\\
&\mathfrak{B}^{1/2,1/2}=-\frac{g^2 M^2 (s-R_p)}{32\pi F^2}\ \mbox{, }\\
&\mathfrak{C}^{1/2,1/2}=\frac{M(s-R_m)}{32F^2\pi}-\frac{Mg^2(sc_R-M^2R_m)}{32F^2\pi(s-M^2)}\ \mbox{, }\\
&\mathfrak{D}^{1/2,1/2}=-\frac{g^2 M^3 (s-R_{m})}{32 \pi  F^2}\ \mbox{. }
\end{align}
\item{$I=3/2, J=1/2$}
\begin{align}
&\mathfrak{A}^{3/2,1/2}=\frac{-(s-R_p)+g^2(s+3M^2-m^2)}{64F^2\pi}\ \mbox{, }\\
&\mathfrak{B}^{3/2,1/2}=\frac{g^2 M^2 (s-R_p)}{16\pi F^2}\ \mbox{, }\\
&\mathfrak{C}^{3/2,1/2}=\frac{M\big[-(s-R_m)+g^2(3s+R_m)\big]}{64\pi F^2}\ \mbox{, }\\
&\mathfrak{D}^{3/2,1/2}=\frac{g^2 M^3 (s-R_m)}{16 \pi  F^2}\ \mbox{. }
\end{align}
\item{$I=1/2, J=3/2$}
\begin{align}
&\mathfrak{A}^{1/2,3/2}=0\ \mbox{, }\\
&\mathfrak{B}^{1/2,3/2}=-\frac{g^2 M^2 (s-R_p)}{32\pi F^2}\ \mbox{, }\\
&\mathfrak{C}^{1/2,3/2}=0\ \mbox{, }\\
&\mathfrak{D}^{1/2,3/2}=-\frac{g^2 M^3 (s-R_m)}{32\pi F^2}\ \mbox{. }
\end{align}
\item{$I=3/2, J=3/2$}
\begin{align}
&\mathfrak{A}^{3/2,3/2}=0\ \mbox{, }\\
&\mathfrak{B}^{3/2,3/2}=\frac{g^2 M^2 (s-R_p)}{16\pi F^2}\ \mbox{, }\\
&\mathfrak{C}^{3/2,3/2}=0\ \mbox{, }\\
&\mathfrak{D}^{3/2,3/2}=\frac{g^2 M^3 (s-R_m)}{16\pi F^2}\ \mbox{. }
\end{align}
\end{itemize}
\subsection{$\mathcal{O}(p^2)$ amplitudes}
The helicity amplitudes of $\mathcal{O}(p^2)$ with $J=1/2$ are
\begin{align}
&T_{++}^{I=1/2,J=1/2}=\frac{c_2 I_{C1}^{1/2}-8M^2\big\{4c_1 m^2-c_3 I_{Ct}^{1/2}+c_4\big[I_{C1}^{1/2}-2(s-R_p)\big]\big\}}
{128\pi M F^2}\ \mbox{, }\\
&T_{+-}^{I=1/2,J=1/2}=\frac{32c_4 M^4(s-R_m)
+(s+R_m)\big[c_2 I_{S2}^{1/2}+8M^2(c_3 I_{St}^{1/2}-c_4 I_{S1}^{1/2}-4c_1 m^2)\big]}
{256\pi M^2 F^2\sqrt{s}}\ \mbox{, }\\
&T_{++}^{I=3/2,J=1/2}=\frac{c_2 I_{C2}^{1/2}+4M^2\big\{-8c_1 m^2+2c_3 I_{Ct}^{1/2}+c_4\big[I_{C1}^{1/2}-2(s-R_p)\big]\big\}}
{128\pi M F^2}\ \mbox{, }\\
&T_{+-}^{I=3/2,J=1/2}=\frac{-16c_4 M^4(s-R_m)
+(s+R_m)\big[c_2 I_{S2}^{1/2}+4M^2(2c_3 I_{St}^{1/2}+c_4 I_{S1}^{1/2}-8c_1 m^2)\big]}
{256\pi M^2 F^2\sqrt{s}}\ \mbox{, }
\end{align}
where the $I_{\cdots}^{1/2}$ are some partial-wave integrals, specifically,
\begin{align}
&I_{C1}^{1/2}=\int_{-1}^{1}\frac{z_s+1}{2}\big[s-u(s,z_s)\big]=-\frac{R_m^2}{3s}-\frac{4}{3}R_p+\frac{5}{3}s\ \mbox{, }\\
&I_{C2}^{1/2}=\int_{-1}^{1}\frac{z_s+1}{2}\big[s-u(s,z_s)\big]^2=\frac{R_m^4}{6s^2}+\frac{2R_m^2R_p}{3s}+(m^4+6m^2M^2+M^4)-\frac{14}{3}R_ps+\frac{17s^2}{6}\ \mbox{, }\\
&I_{Ct}^{1/2}=\int_{-1}^{1}\frac{z_s+1}{2}\big[2m^2-t(s,z_s)\big]=\frac{R_m^2}{3s}+\frac{2}{3}(2m^2-M^2)+\frac{s}{3}\ \mbox{, }
\end{align}
and
\begin{align}
&I_{S1}^{1/2}=\int_{-1}^{1}\frac{z_s-1}{2}\big[s-u(s,z_s)\big]=-\frac{2R_m^2}{3s}-\frac{2}{3}R_p+\frac{4}{3}s\ \mbox{, }\\
&I_{S2}^{1/2}=\int_{-1}^{1}\frac{z_s-1}{2}\big[s-u(s,z_s)\big]^2=\frac{R_m^4}{2s^2}+\frac{2R_m^2R_p}{3s}-\frac{3m^4-14m^2M^2+3M^4}{3}-2R_ps+\frac{11s^2}{6}\ \mbox{, }\\
&I_{St}^{1/2}=\int_{-1}^{1}\frac{z_s-1}{2}\big[2m^2-t(s,z_s)\big]=\frac{2R_m^2}{3s}+\frac{2}{3}(m^2-2M^2)+\frac{2s}{3}\ \mbox{, }
\end{align}
with
\begin{equation}
t(s,z_s)=2m^2-\frac{s^2-R_m^2}{2s}-\frac{(s-s_L)(s-s_R)}{2s}z_s\ \mbox{. }
\end{equation}
For the case of $J=3/2$,
\begin{align}
&T_{++}^{I=1/2,J=3/2}=\frac{c_2 I_{C2}^{3/2}+8M^2(c_3 I_{Ct}^{3/2}-c_4I_{C1}^{3/2})}{128\pi M F^2}\ \mbox{, }\\
&T_{+-}^{I=1/2,J=3/2}=\frac{(s+R_m)\big[c_2 I_{S2}^{3/2}+8M^2(c_3 I_{St}^{3/2}-c_4 I_{S1}^{3/2})\big]}{256\pi M^2 F^2\sqrt{s}}\ \mbox{, }\\
&T_{++}^{I=3/2,J=3/2}=\frac{c_2 I_{C2}^{3/2}+4M^2(2c_3 I_{Ct}^{3/2}+c_4I_{C1}^{3/2})}{128\pi M F^2}\ \mbox{, }\\
&T_{+-}^{I=3/2,J=3/2}=\frac{(s+R_m)\big[c_2 I_{S2}^{3/2}+4M^2(2c_3 I_{St}^{3/2}+c_4 I_{S1}^{3/2})\big]}{256\pi M^2 F^2\sqrt{s}}\ \mbox{, }
\end{align}
with
\begin{align}
&I_{C1}^{3/2}=\int_{-1}^{1}\frac{(z_s+1)(3z_s-1)}{4}\big[s-u(s,z_s)\big]=\frac{-2m^2(s+M^2)+m^4+(s-M^2)^2}{6s}\ \mbox{, }\\
&I_{C2}^{3/2}=\int_{-1}^{1}\frac{(z_s+1)(3z_s-1)}{4}\big[s-u(s,z_s)\big]^2=\frac{2(s-s_L)(s-s_R)(4s^2-3R_ps-R_m^2)}{15s^2}\ \mbox{, }\\
&I_{Ct}^{3/2}=\int_{-1}^{1}\frac{(z_s+1)(3z_s-1)}{4}\big[2m^2-t(s,z_s)\big]=-\frac{-2m^2(s+M^2)+m^4+R_m^2}{6s}\ \mbox{, }
\end{align}
and
\begin{align}
&I_{S1}^{3/2}=\int_{-1}^{1}\frac{(z_s-1)(3z_s+1)}{4}\big[s-u(s,z_s)\big]=\frac{-2m^2(s+M^2)+m^4+(s-M^2)^2}{6s}\ \mbox{, }\\
&I_{S2}^{3/2}=\int_{-1}^{1}\frac{(z_s-1)(3z_s+1)}{4}\big[s-u(s,z_s)\big]^2=\frac{(s-s_L)(s-s_R)(7s^2-4R_ps-3R_m^2)}{15s^2}\ \mbox{, }\\
&I_{St}^{3/2}=\int_{-1}^{1}\frac{(z_s-1)(3z_s+1)}{4}\big[2m^2-t(s,z_s)\big]=-\frac{-2m^2(s+M^2)+m^4+(s-M^2)^2}{6s}\ \mbox{. }
\end{align}
\section{Examinations on the uncertainties of left-hand cut integrals}\label{app:uncer}
\subsection{Varying low energy constants}\label{app:ci}
\begin{figure}[htbp]
\center
\subfigure[]{
\label{ci:subfig:S11}
\scalebox{1.2}[1.2]{\includegraphics[width=0.4\textwidth]{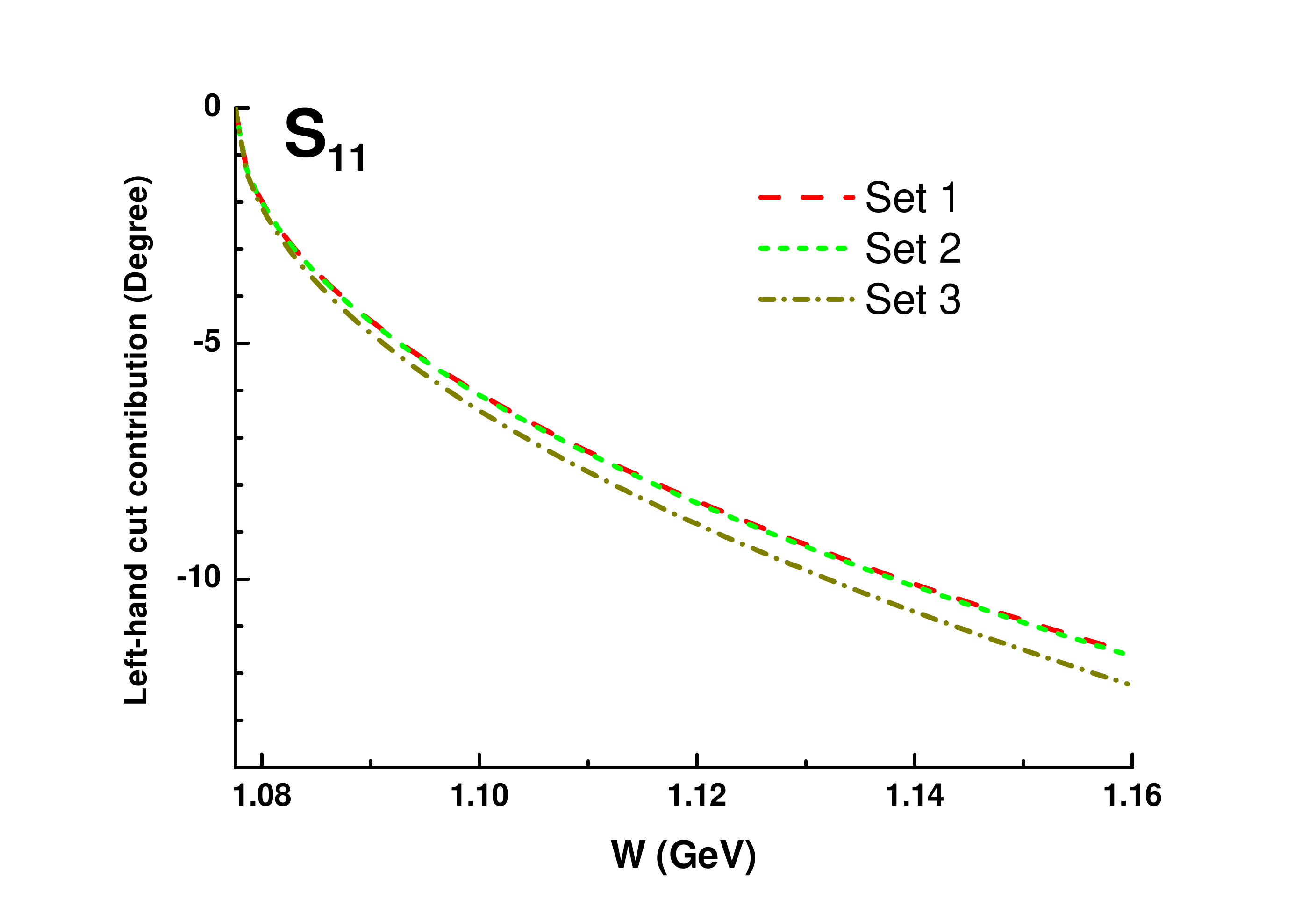}}}
\subfigure[]{
\label{ci:subfig:S31}
\scalebox{1.2}[1.2]{\includegraphics[width=0.4\textwidth]{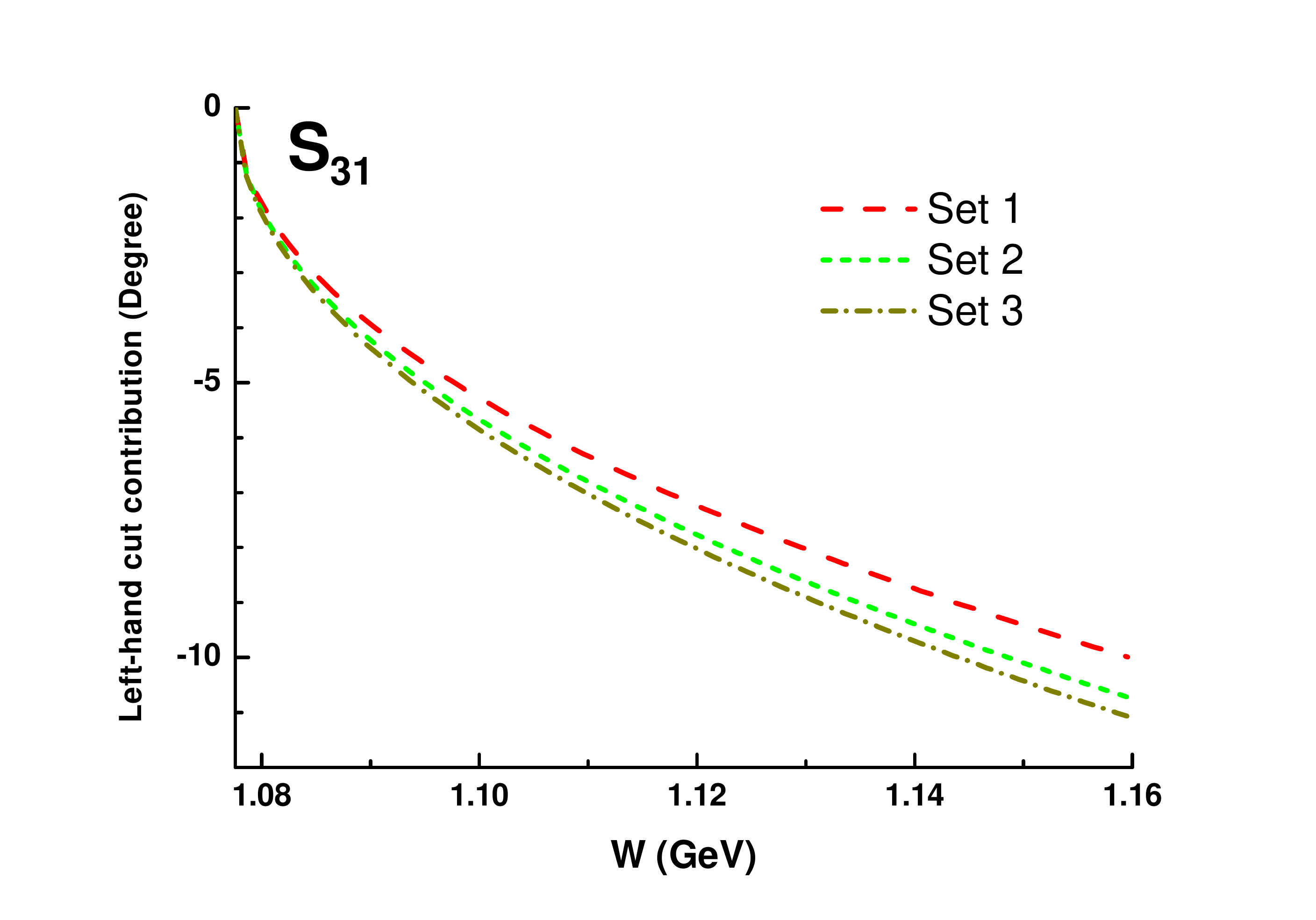}}}
\subfigure[]{
\label{ci:subfig:P11}
\scalebox{1.2}[1.2]{\includegraphics[width=0.4\textwidth]{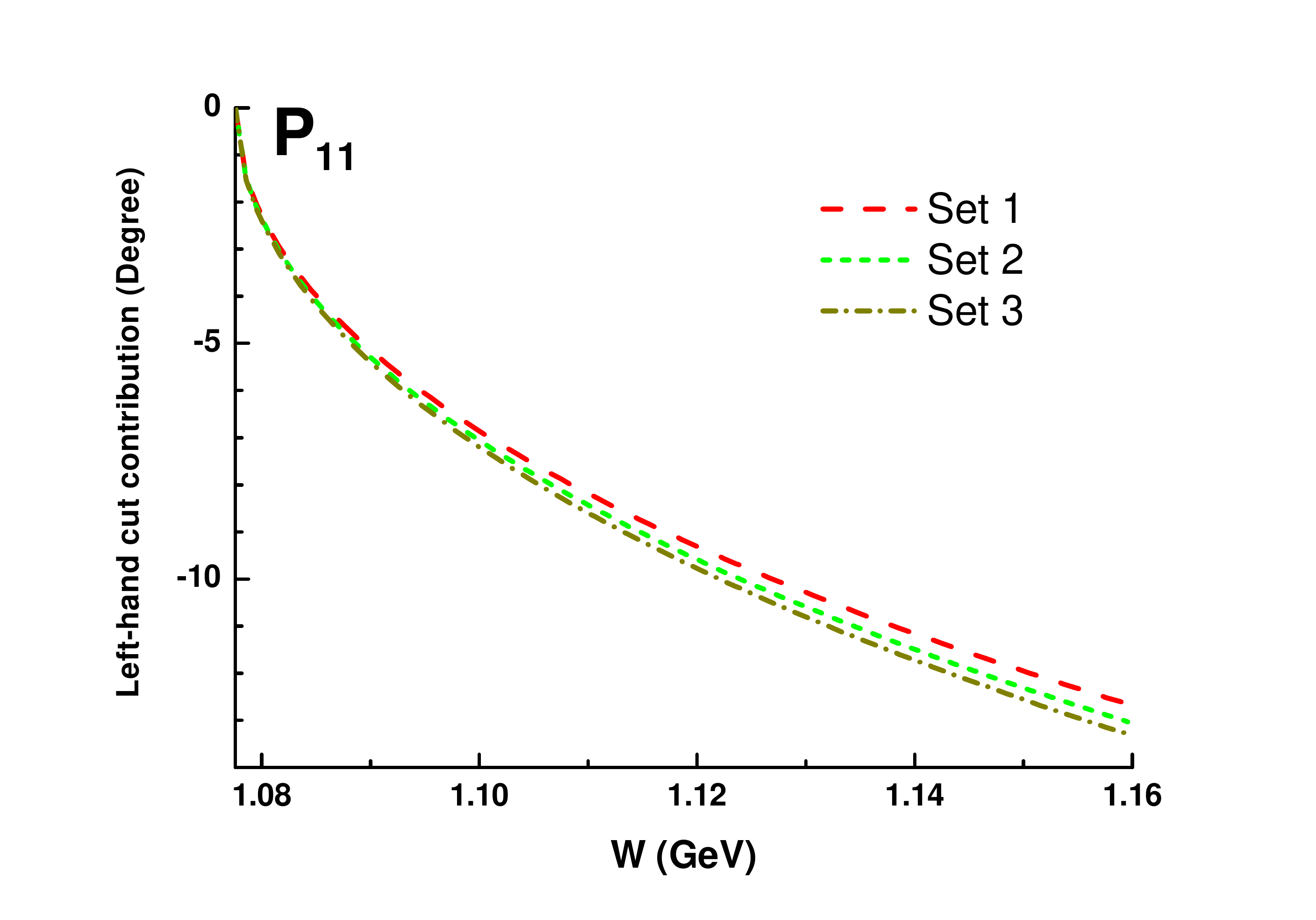}}}
\subfigure[]{
\label{ci:subfig:P31}
\scalebox{1.2}[1.2]{\includegraphics[width=0.4\textwidth]{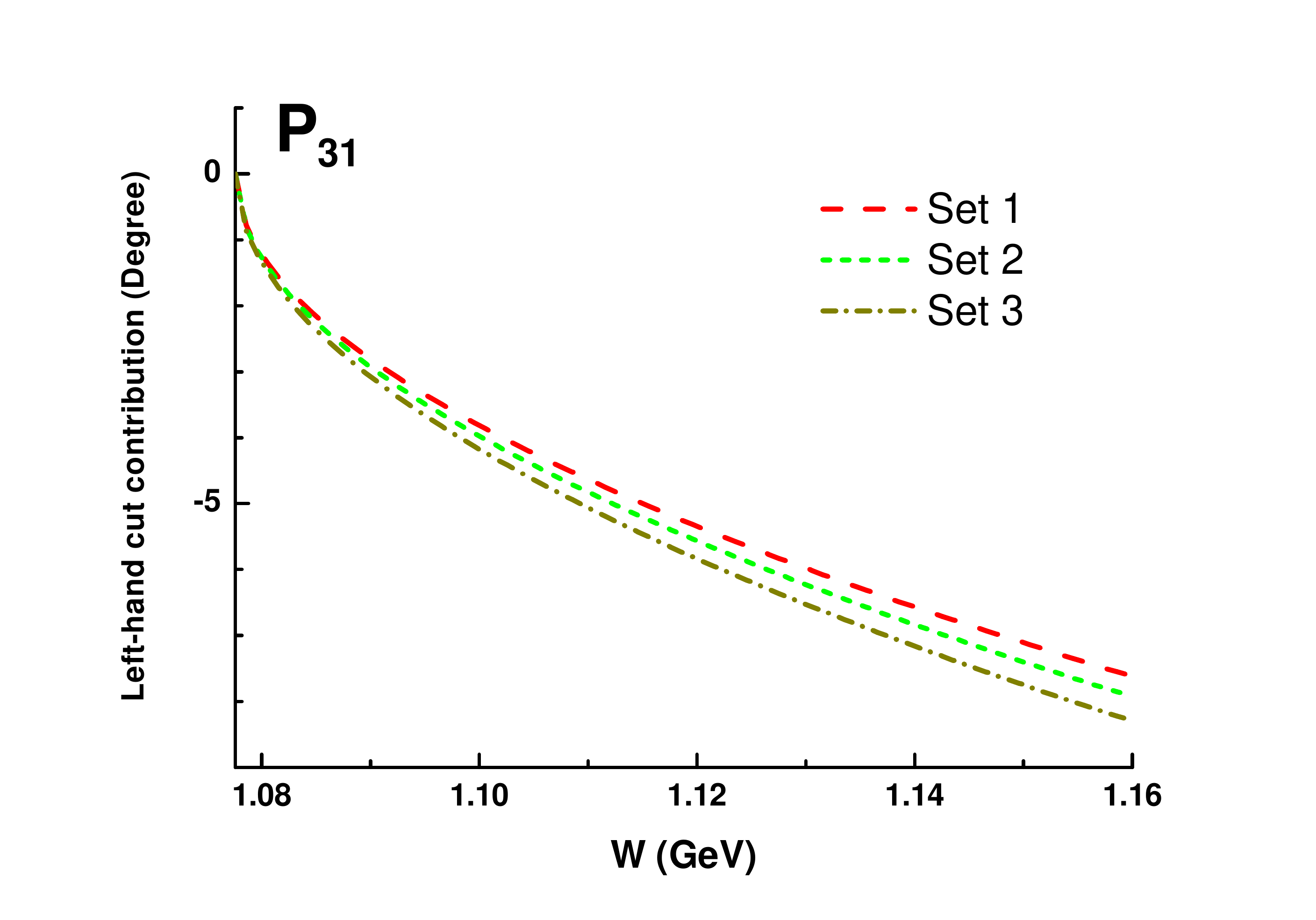}}}
\subfigure[]{
\label{ci:subfig:P13}
\scalebox{1.2}[1.2]{\includegraphics[width=0.4\textwidth]{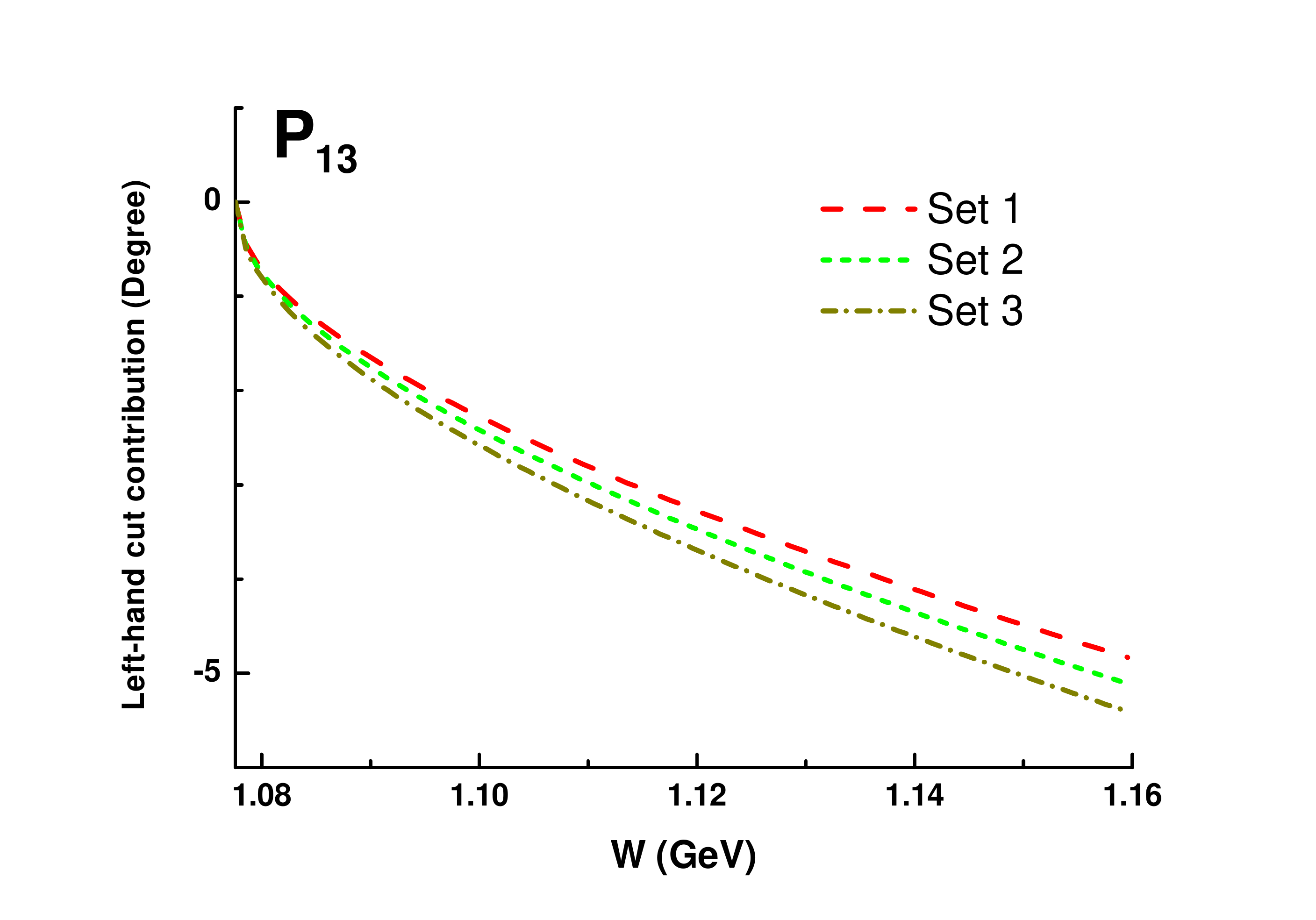}}}
\subfigure[]{
\label{ci:subfig:P33}
\scalebox{1.2}[1.2]{\includegraphics[width=0.4\textwidth]{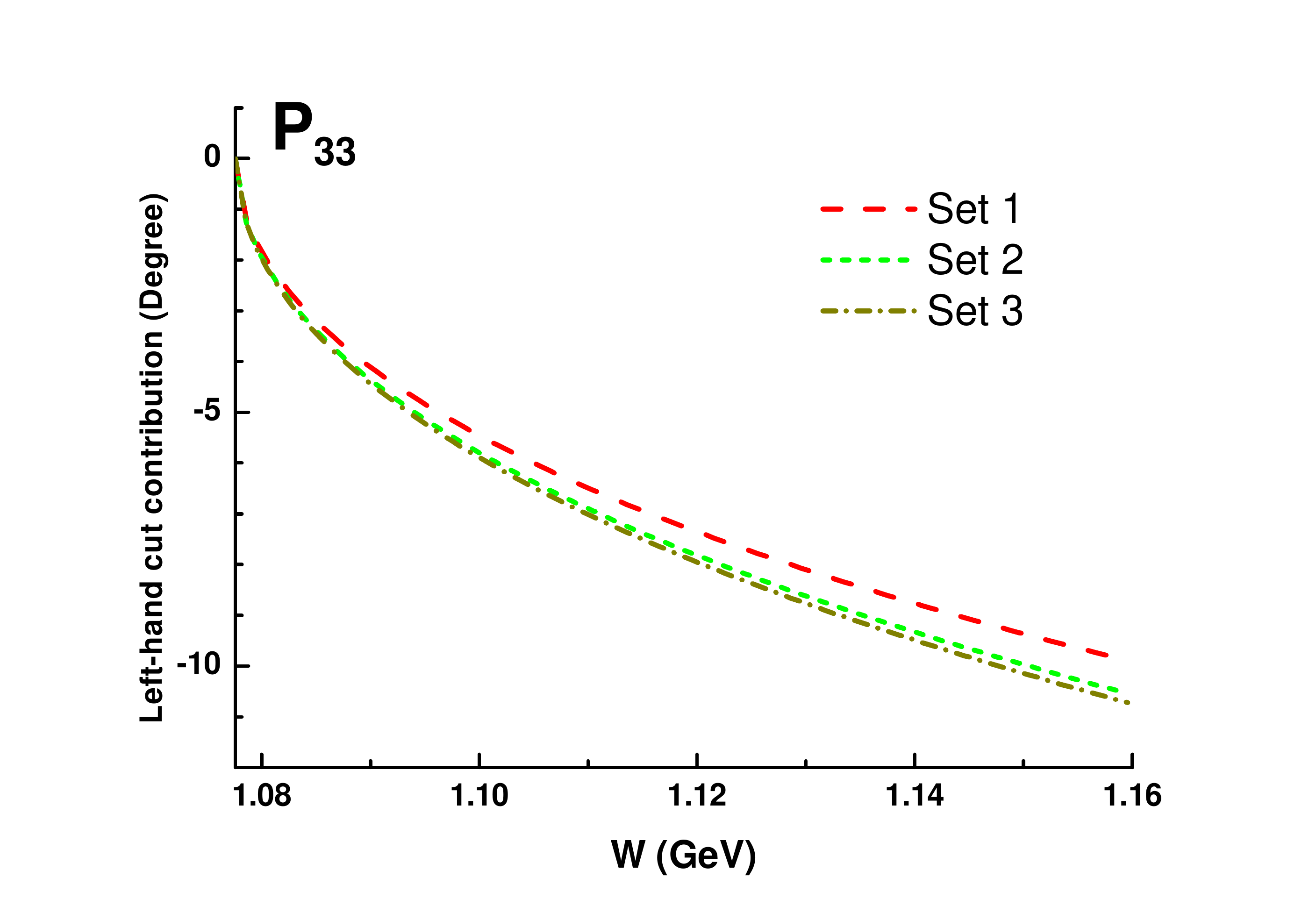}}}
\caption{Left-hand cut contributions with different values of low energy constants $c_i$. }\label{fig:ci}
\end{figure}

In this section we show that the variation of $\mathcal{O}(p^2)$ low energy constants $c_i$ in Eq.~\eqref{p2la} has little impact on the results of PKU representation analyses. Three sets of $c_i$ values are employed: set 1 is the result of $K$-matrix fit to the phase shift data in this paper, see Section \ref{res}; set 2 is from the $K$-matrix fit to the phase shift data at $\mathcal{O}(p^3)$ level, see Table.~1 (Fit II) in Ref.~\cite{Chen:2012nx}; set 3 is the values that are given by the fit to the subthreshold parameters from Roy-Stainer analyses at $\mathcal{O}(p^2)$ level, see Table.~1 in Ref.~\cite{Siemens:2016jwj}. From Fig.~\ref{fig:ci} it is clear that the left-hand cut contributions of different $c_i$ values are quite close to each other, hence the main results of this paper are irrelevant to different $c_i$ determinations.
\subsection{Changing integration cut-off $s_{c}$}\label{app:cutoff}
\begin{figure}[htbp]
\center
\subfigure[]{
\label{cutoff:subfig:S11}
\scalebox{1.2}[1.2]{\includegraphics[width=0.4\textwidth]{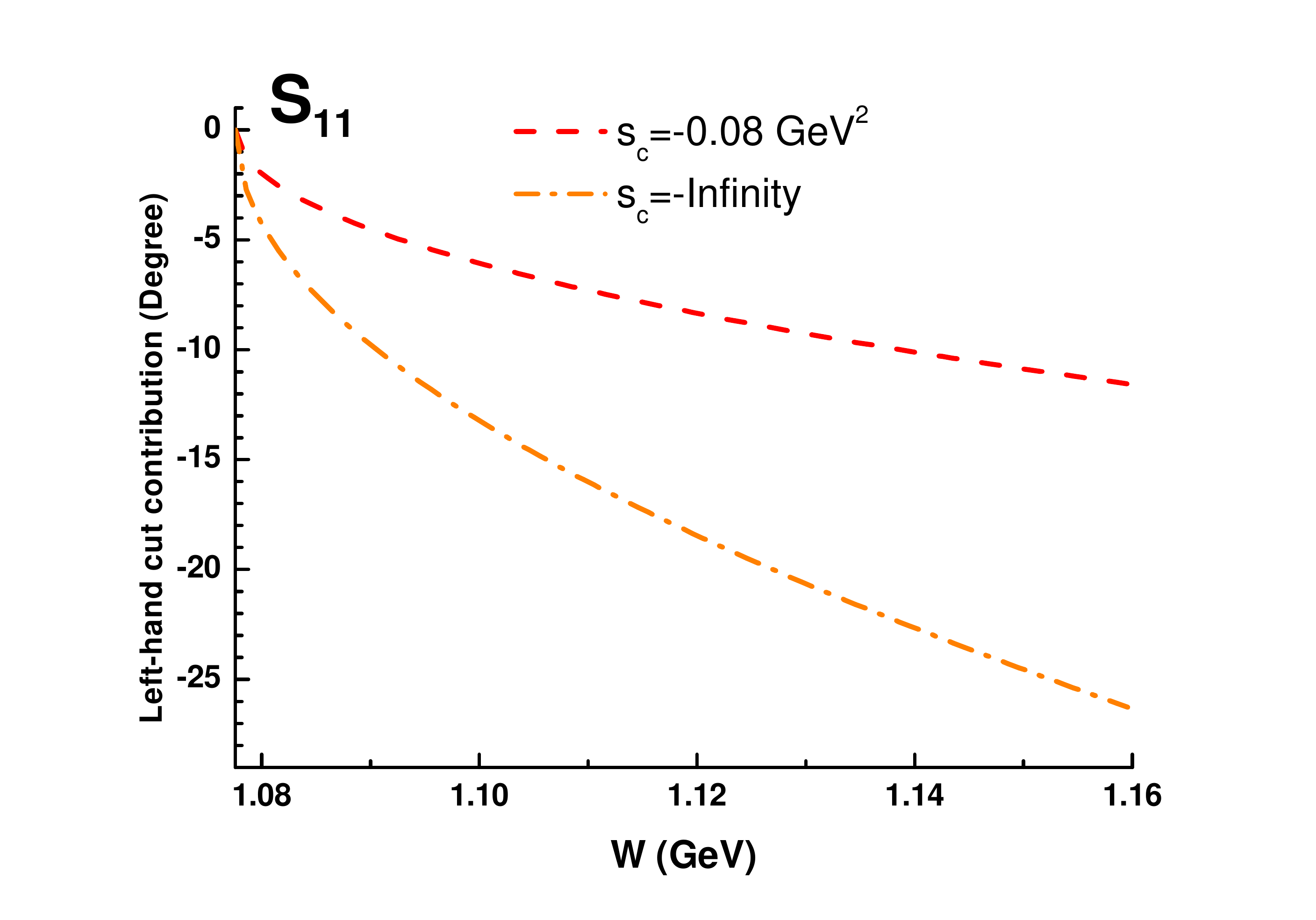}}}
\subfigure[]{
\label{cutoff:subfig:S31}
\scalebox{1.2}[1.2]{\includegraphics[width=0.4\textwidth]{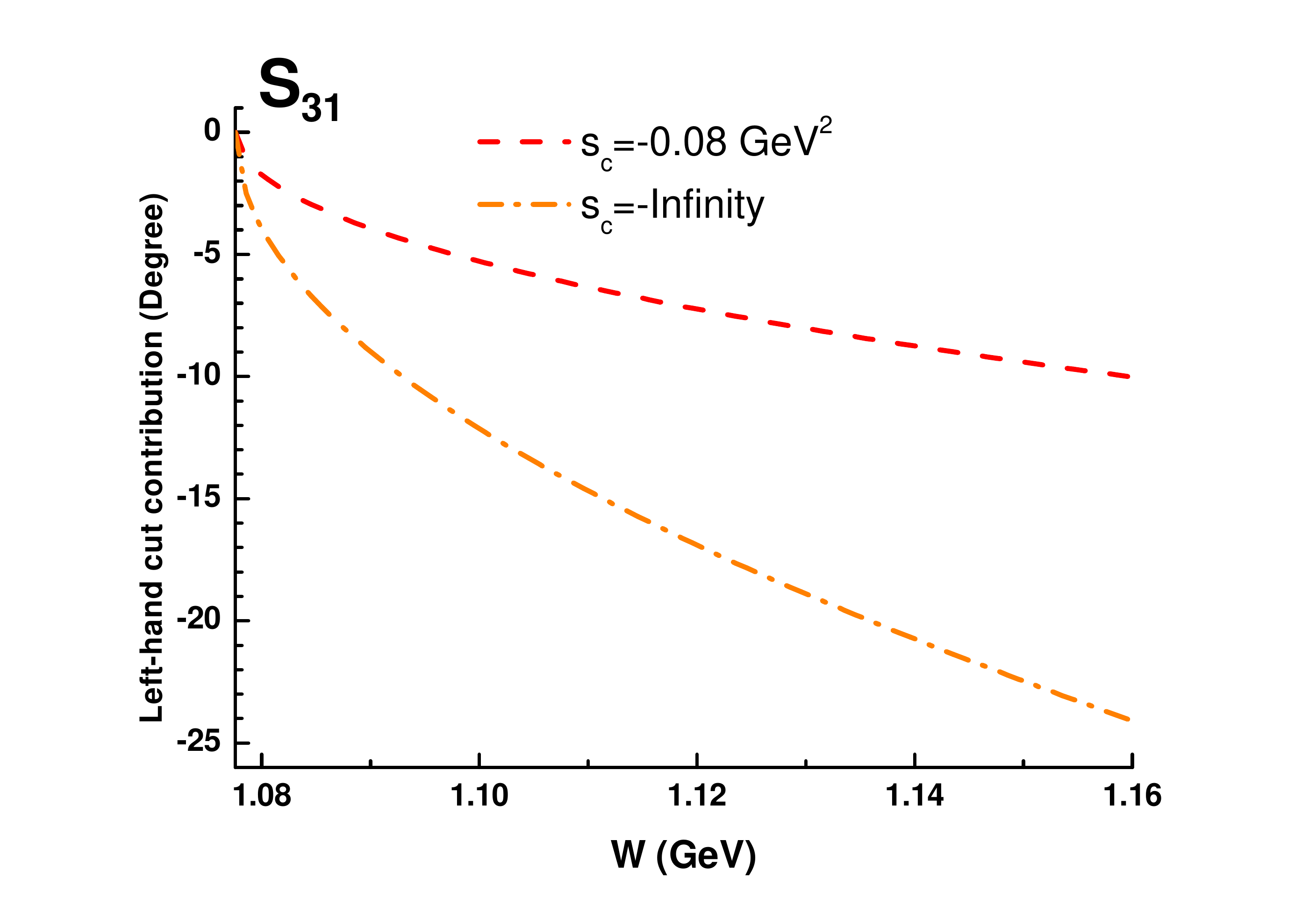}}}
\subfigure[]{
\label{cutoff:subfig:P11}
\scalebox{1.2}[1.2]{\includegraphics[width=0.4\textwidth]{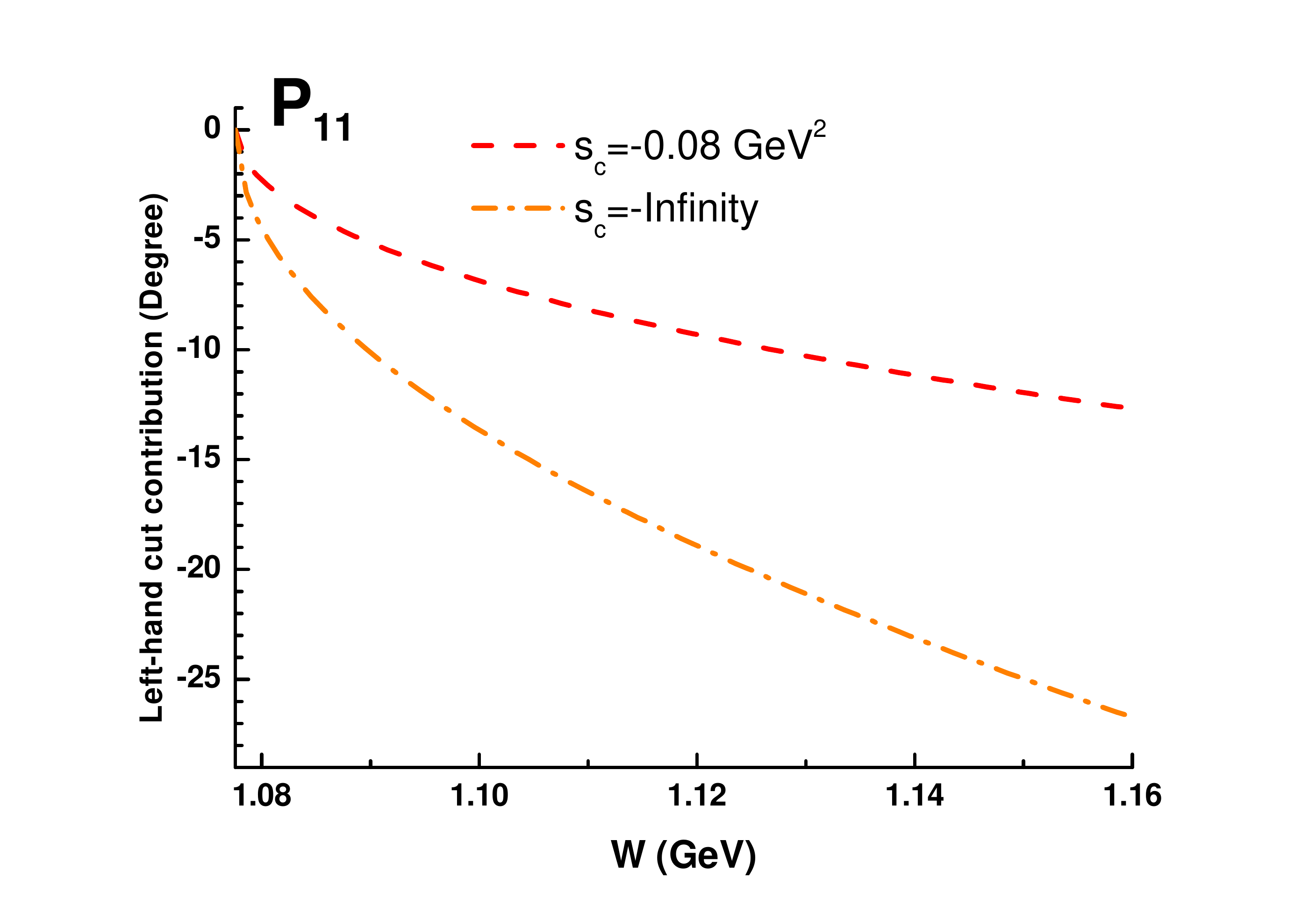}}}
\subfigure[]{
\label{cutoff:subfig:P31}
\scalebox{1.2}[1.2]{\includegraphics[width=0.4\textwidth]{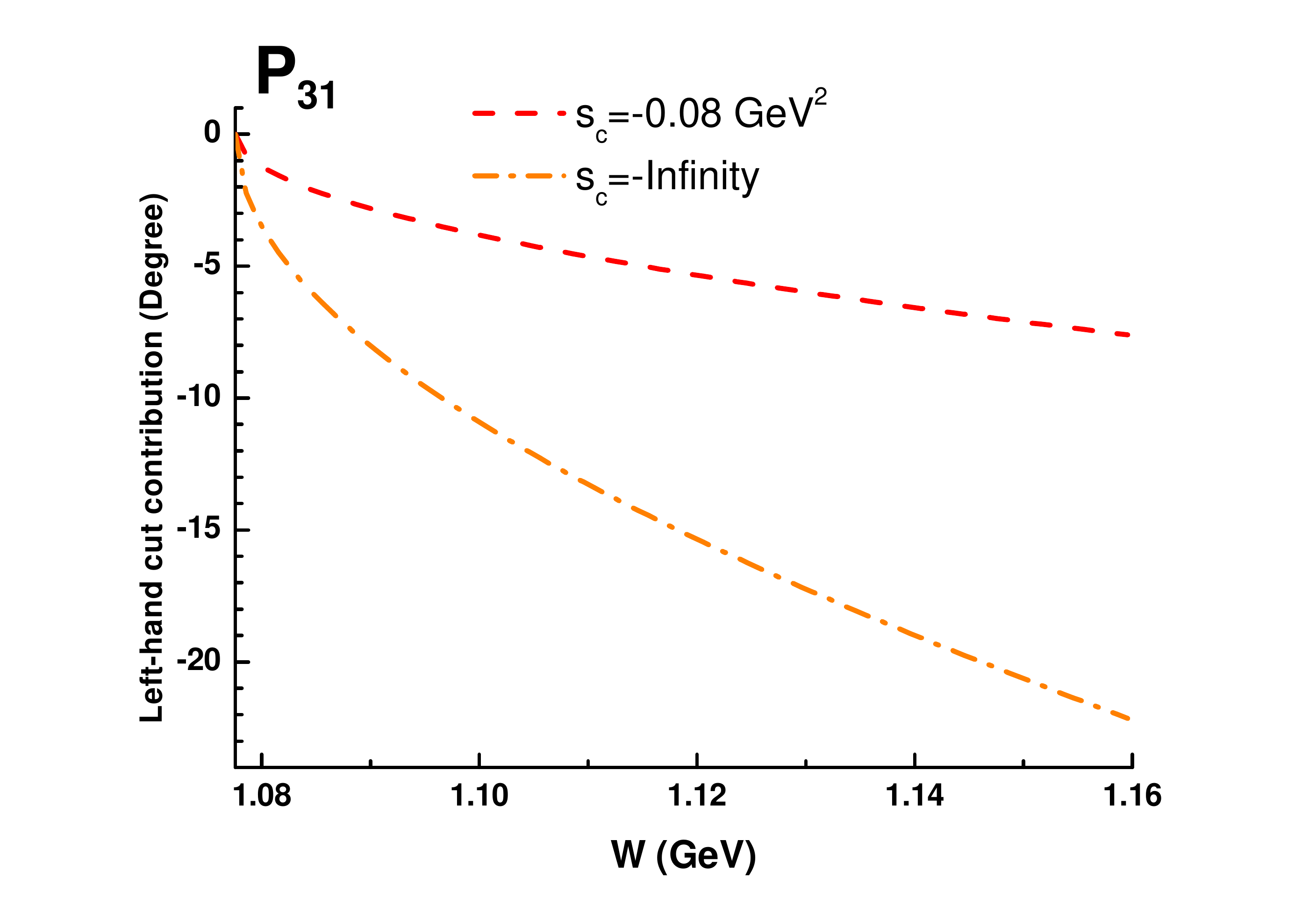}}}
\subfigure[]{
\label{cutoff:subfig:P13}
\scalebox{1.2}[1.2]{\includegraphics[width=0.4\textwidth]{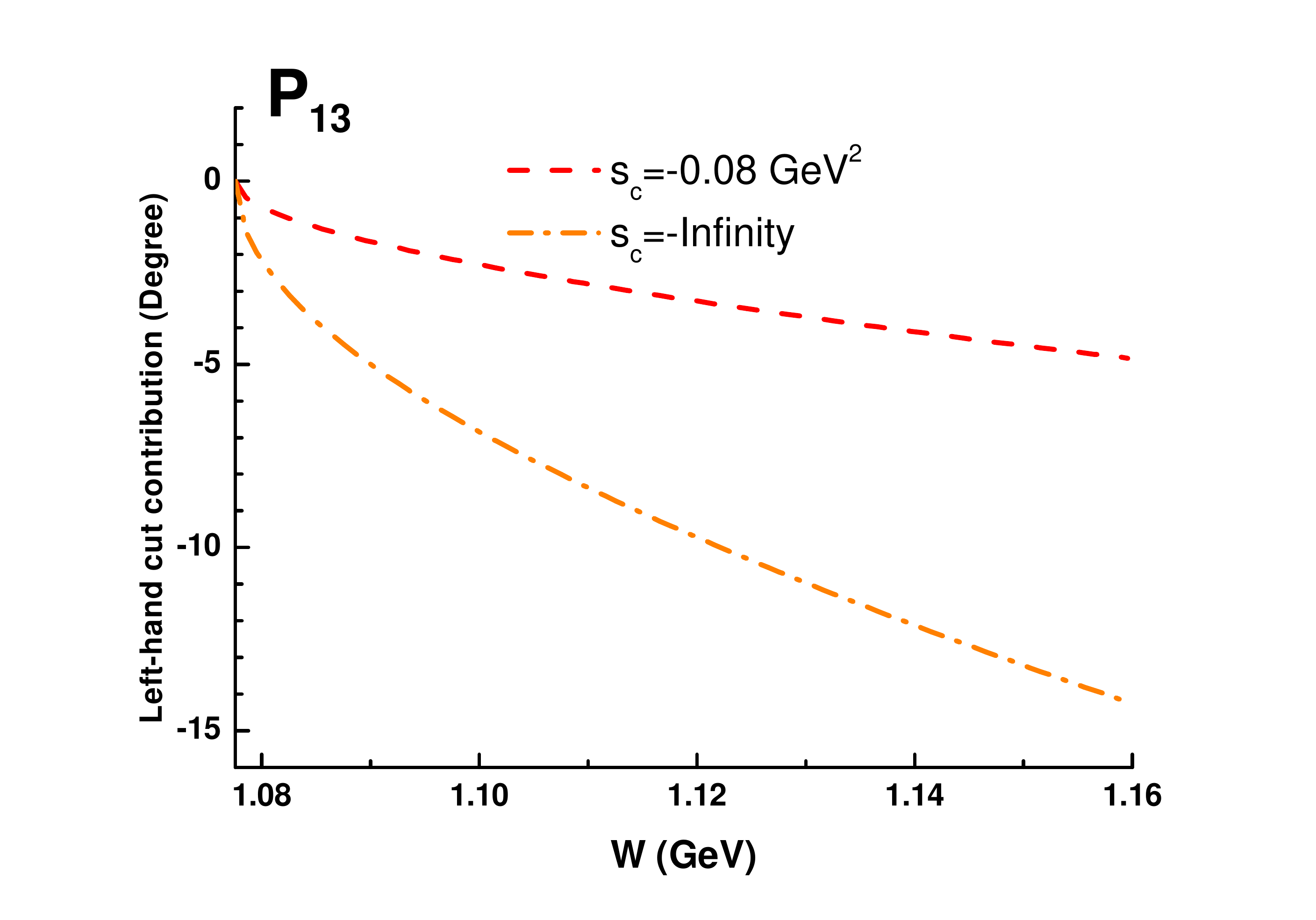}}}
\subfigure[]{
\label{cutoff:subfig:P33}
\scalebox{1.2}[1.2]{\includegraphics[width=0.4\textwidth]{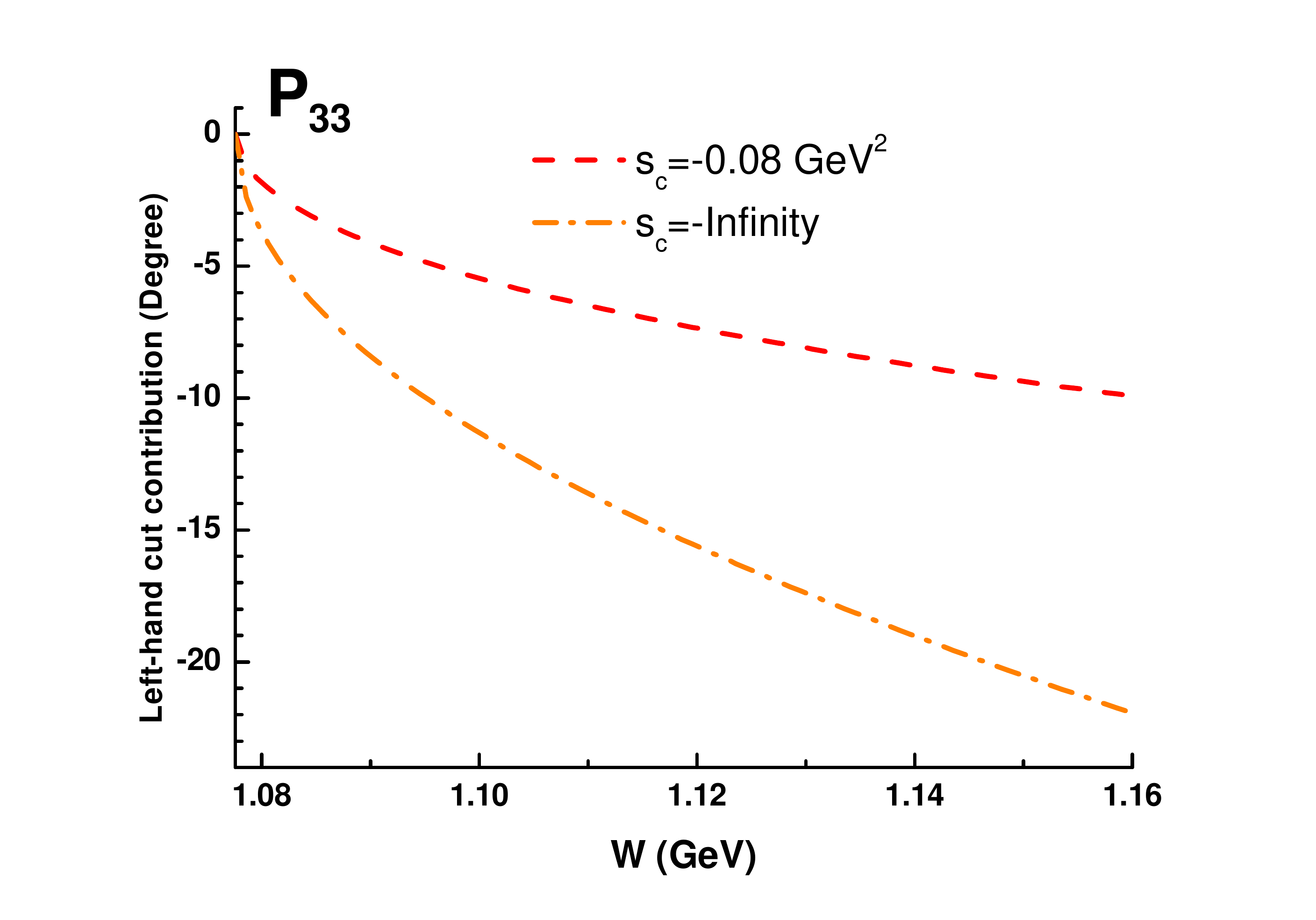}}}
\caption{Left-hand cut contributions given by PKU representation under different cut-off values. }\label{fig:cutoff}
\end{figure}

In what follows we investigate the effect of the variations of the cut-off $s_{c}$ in Eq.~\eqref{fsint}: $s_{c}$ ranges from $-0.08$ GeV$^2$ to $-\infty$, see Fig.~\ref{fig:cutoff}. It is found that the cut-off parameter is actually of some numerical importance, but the qualitative picture never changes. Besides, the missing contributions in $S_{11}$ and $P_{11}$ channels can never be blotted out by changing $s_{c}$.

In $S_{11}$ channel, it is found that the extra pole stays below threshold when $s_{c}$ changes, see Table.~\ref{tab:S11sh}.
\begin{table}[htbp]
\begin{center}
 \begin{tabular}  {| c | c | c ||}
  \hline
  $s_{c}$ (GeV$^2$)  & Pole position (GeV) & Fit quality $\chi^2/\text{d.o.f}$\\
  \hline
  $-0.08$ & $0.808-0.055i$ & $0.109$\\
  \hline
  $-1$ &$0.822-0.139i$ & $0.076$\\
  \hline
  $-9$ &$0.883-0.195i$ & $0.034$\\
  \hline
  $\infty$ &$0.914-0.205i$ & $0.018$\\
  \hline
 \end{tabular}\\
 \caption{The $S_{11}$ hidden pole fit with different choices of $s_{c}$. }\label{tab:S11sh}
\end{center}
\end{table}
\subsection{Truncating chiral orders: $\mathcal{O}(p^1)$ versus $\mathcal{O}(p^2)$}\label{app:p1}
\begin{figure}[tbp]
\center
\subfigure[]{
\label{p12:subfig:S11p1}
\scalebox{1.0}[1.0]{\includegraphics[width=0.4\textwidth]{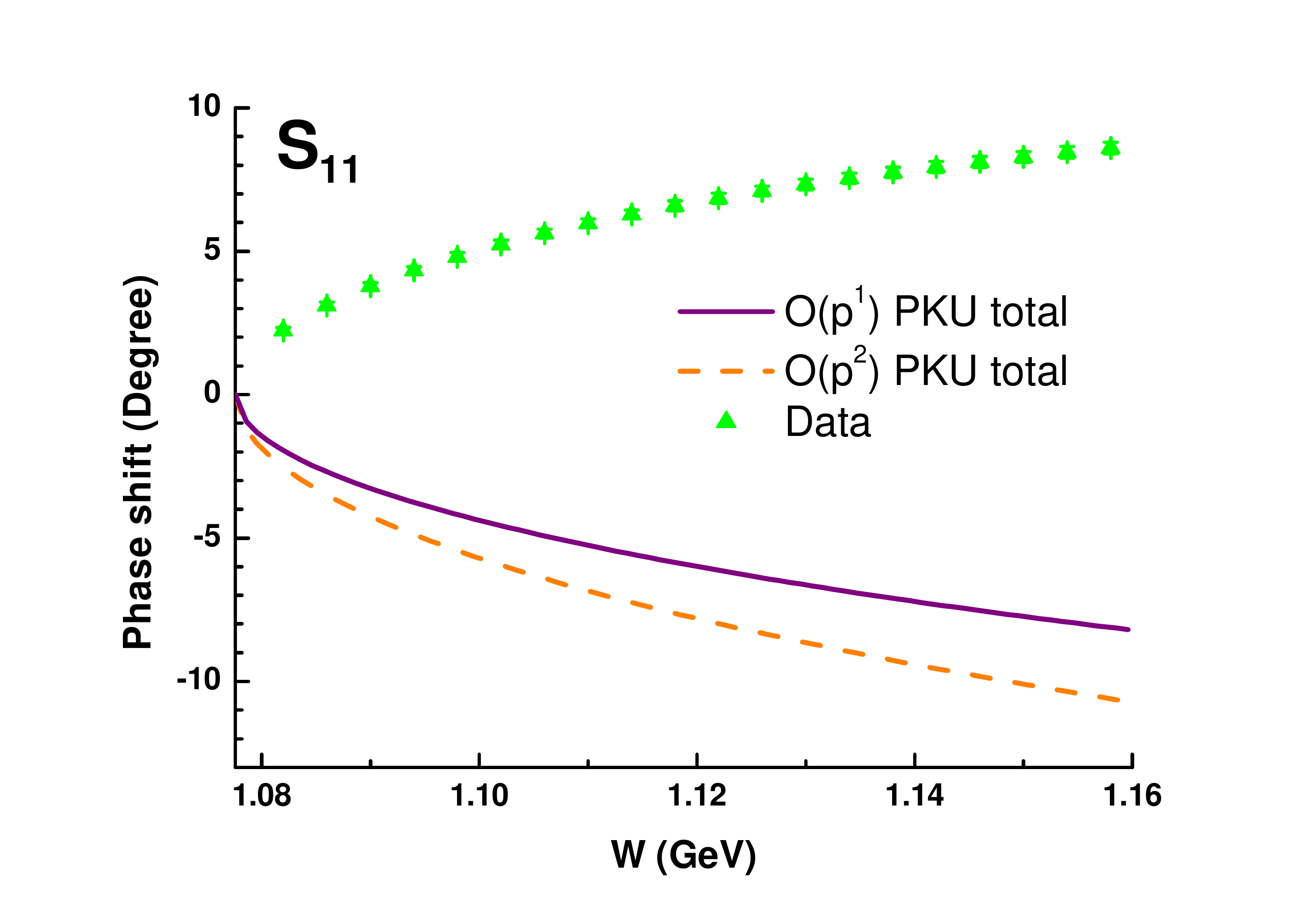}}}
\subfigure[]{
\label{p12:subfig:S31p1}
\scalebox{1.0}[1.0]{\includegraphics[width=0.4\textwidth]{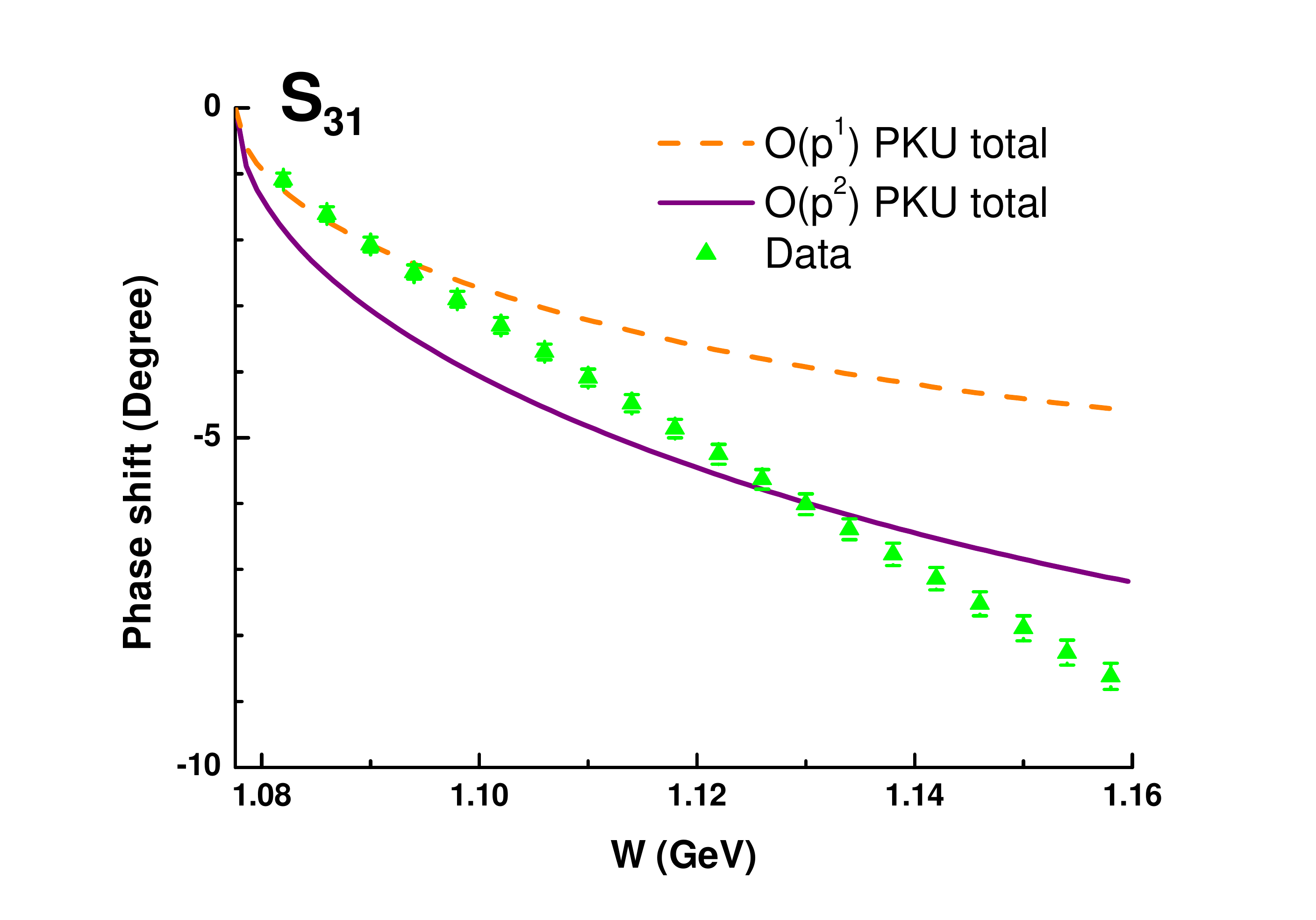}}}
\subfigure[]{
\label{p12:subfig:P11p1}
\scalebox{1.0}[1.0]{\includegraphics[width=0.4\textwidth]{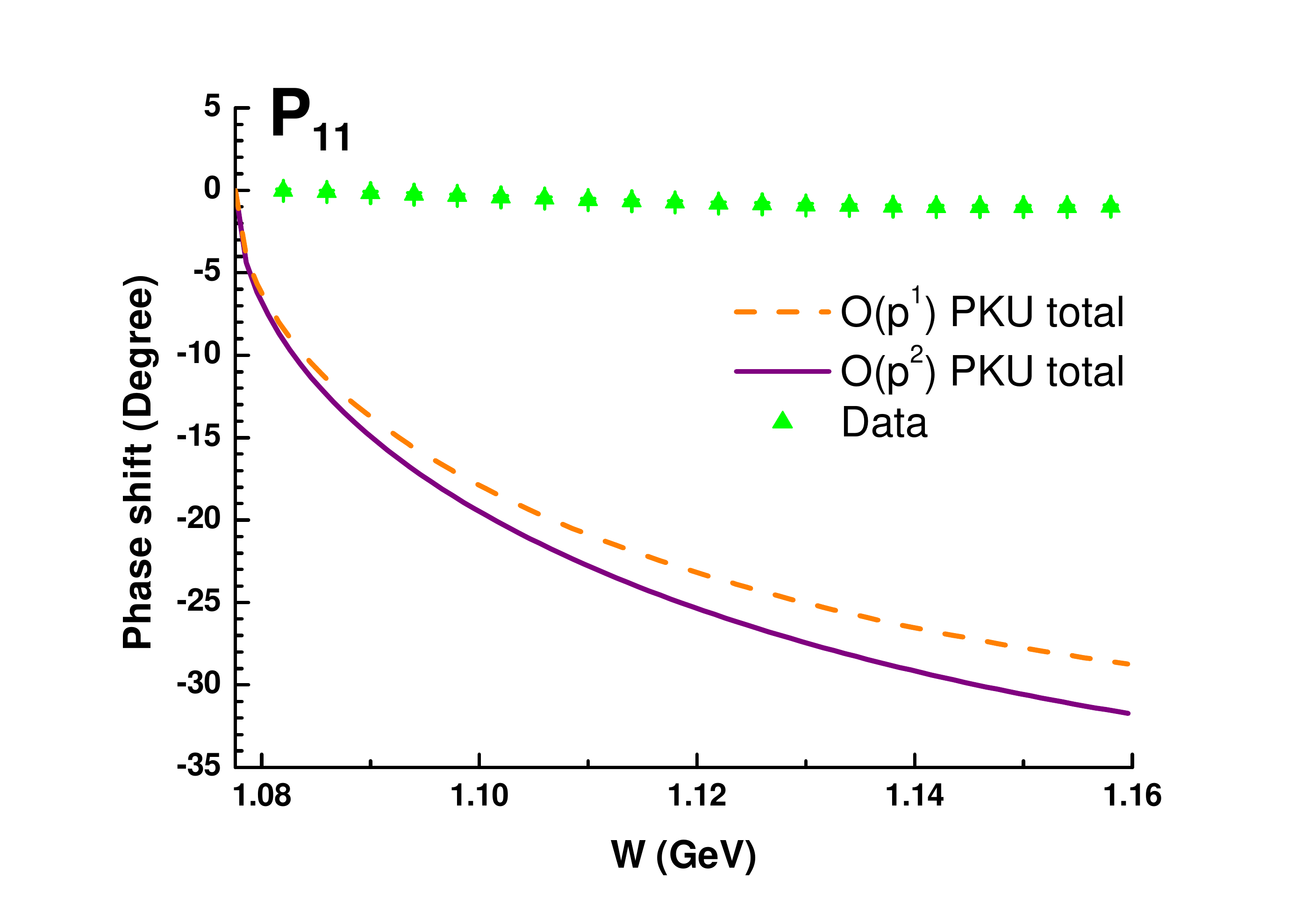}}}
\subfigure[]{
\label{p12:subfig:P31p1}
\scalebox{1.0}[1.0]{\includegraphics[width=0.4\textwidth]{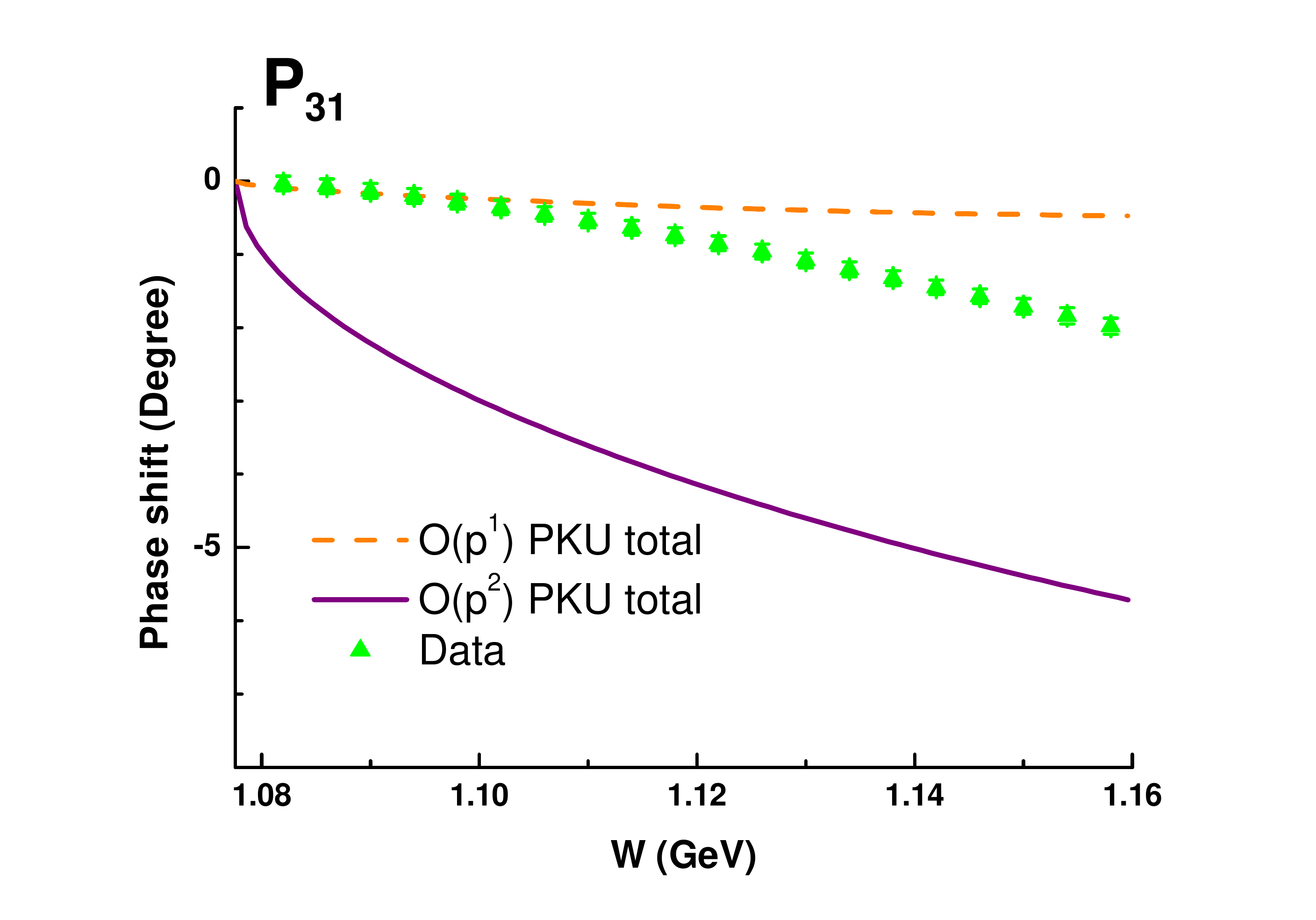}}}
\subfigure[]{
\label{p12:subfig:P13p1}
\scalebox{1.0}[1.0]{\includegraphics[width=0.4\textwidth]{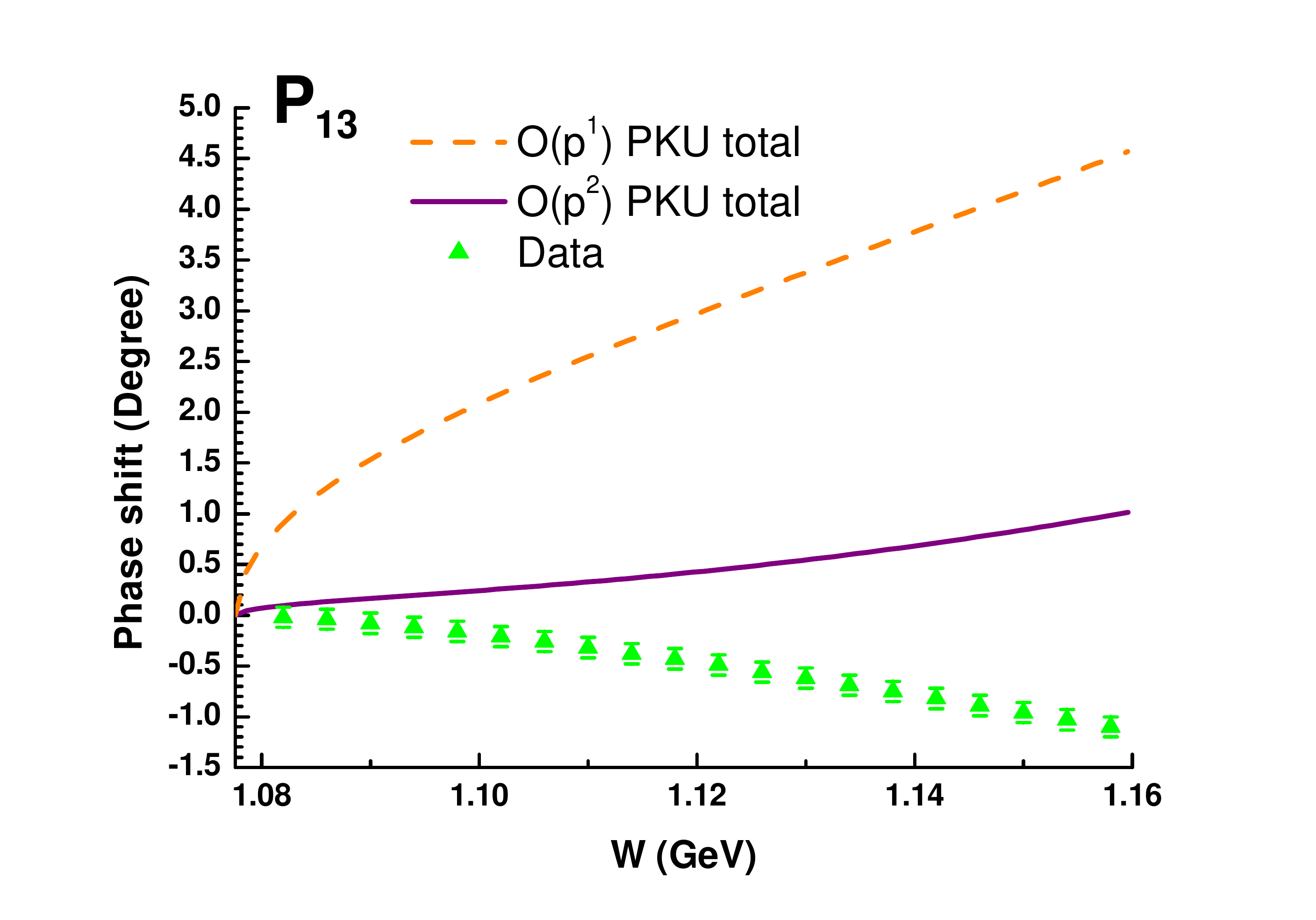}}}
\subfigure[]{
\label{p12:subfig:P33p1}
\scalebox{1.0}[1.0]{\includegraphics[width=0.4\textwidth]{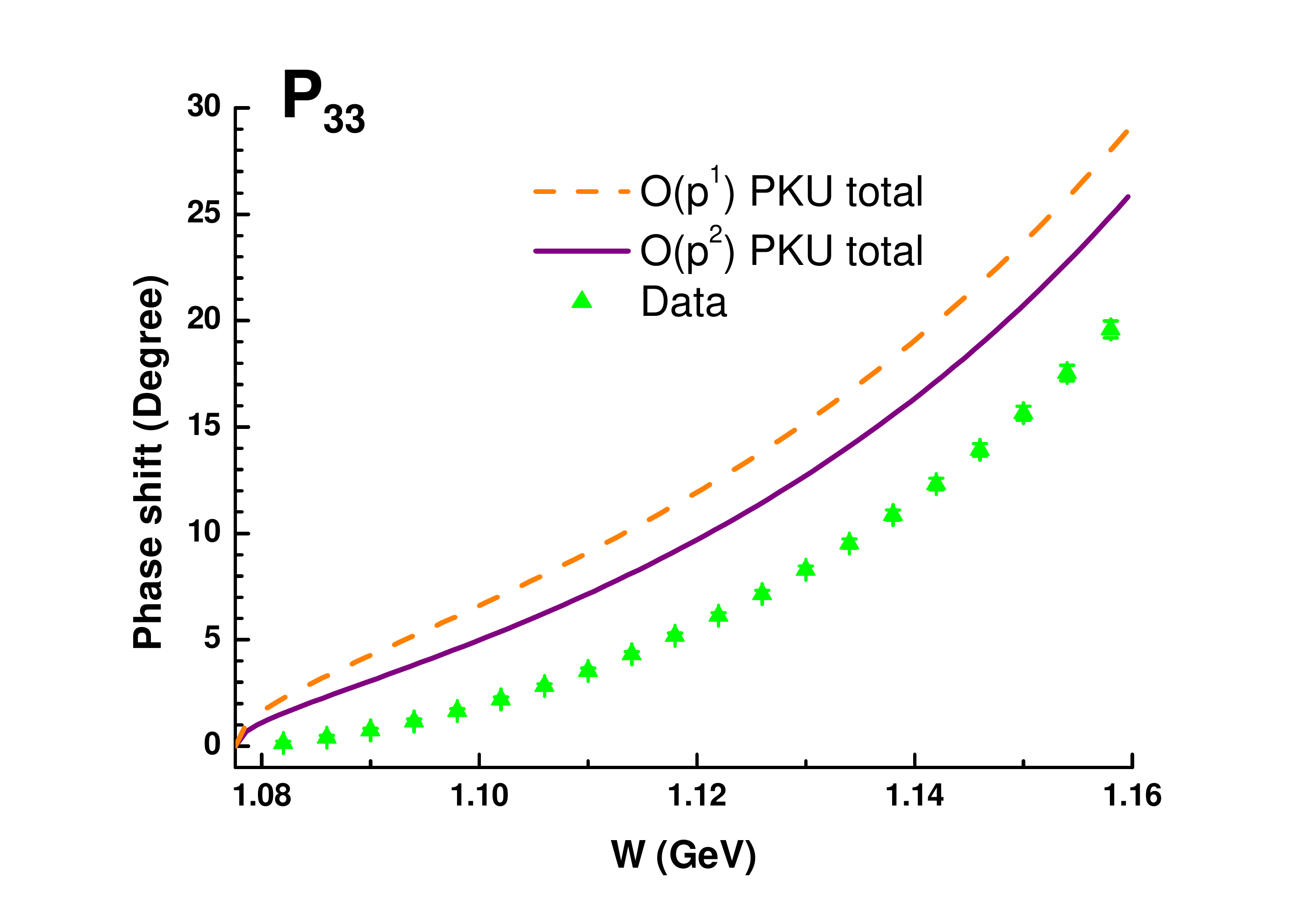}}}
\caption{Comparison of PKU representation analyses between $\mathcal{O}(p^1)$ and $\mathcal{O}(p^2)$. }\label{p1PKU}
\end{figure}

The phase shifts from both $\mathcal{O}(p^1)$ and $\mathcal{O}(p^2)$ chiral amplitudes are plotted in Fig.~\ref{p1PKU}, where the cut-off parameter of the background integral is taken as $-0.08$ GeV$^2$. We present the different estimation of the phase shifts here, for the purpose of comparing the perturbative calculations at different orders. It can be seen that the difference of phase shift results between $\mathcal{O}(p^1)$ and $\mathcal{O}(p^2)$ are only at the order of a few degrees, and at $\mathcal{O}(p^1)$ the $S_{11}$ and $P_{11}$ channels also contain significant disagreements between the known poles plus cut and the data. In general, the qualitative conclusions remain unchanged when the chiral order changes.

\end{document}